\documentclass[twocolumn]{aastex701}

\usepackage{amsmath}
\usepackage{graphicx}   
\usepackage{bm}         
\usepackage{natbib}
\usepackage{etoolbox}
\usepackage{rotating}
\usepackage{array}
\usepackage{booktabs}
\usepackage{amssymb}

\makeatletter
\patchcmd{\NAT@citex}
  {\@citea\NAT@hyper@{%
     \NAT@nmfmt{\NAT@nm}%
\hyper@natlinkbreak{\NAT@aysep\NAT@spacechar}{\@citeb\@extra@b@citeb}%
     \NAT@date}}
  {\@citea\NAT@nmfmt{\NAT@nm}%
   \NAT@aysep\NAT@spacechar\NAT@hyper@{\NAT@date}}{}{}

\patchcmd{\NAT@citex}
  {\@citea\NAT@hyper@{%
     \NAT@nmfmt{\NAT@nm}%
\hyper@natlinkbreak{\NAT@spacechar\NAT@@open\if*#1*\else#1\NAT@spacechar\fi}%
       {\@citeb\@extra@b@citeb}%
     \NAT@date}}
  {\@citea\NAT@nmfmt{\NAT@nm}%
\NAT@spacechar\NAT@@open\if*#1*\else#1\NAT@spacechar\fi\NAT@hyper@{\NAT@date}}
  {}{}
\makeatother

\makeatletter
\DeclareRobustCommand{\textsupsub}[2]{{%
  \m@th\ensuremath{%
    ^{\mbox{\fontsize\sf@size\z@#1}}%
    _{\mbox{\fontsize\sf@size\z@#2}}%
  }%
}}
\makeatother

\newcommand{\lsun}{\mbox{\,$L_\odot$}}
\newcommand{\msun}{\mbox{\,$M_\odot$}}

\newcommand{\kms}{\mbox{\,km\,s$^{-1}$}}

\newcommand{\lbol}{\mbox{$L_{\rm bol}$}}

\newcommand{\ee}[1]{\mbox{${} \times 10^{#1}$}}
\newcommand{\eten}[1]{\mbox{$10^{#1}$}}
\newcommand{\jj}[2]{\mbox{$J = #1\rightarrow#2$}}

\newcommand{\h}{\mbox{$^{\text{h}}$}}
\newcommand{\m}{\mbox{$^{\text{m}}$}}
\newcommand{\s}{\mbox{$^{\text{s}}$}}

\newcommand{\J}{\mbox{$J$}}
\newcommand{\K}{\mbox{$K$}}
\newcommand{\N}{\mbox{$N$}}
\newcommand{\F}{\mbox{$F$}}

\newcommand{\zt}[4]{$M_c=#1$ M$_\odot$; $\Sigma_{\rm cl}=#2$ g cm$^{-2}$; $m_*=#3$ M$_\odot$; $\theta_{\rm view}=#4^\circ$}
\newcommand{\av}{\mbox{$A_V$}}
\newcommand{\chisq}{\mbox{$X^2$}}
\newcommand{\chioned}{\mbox{$X^2_{\rm IMPRO}$}}
\newcommand{\chised}{\mbox{$X^2_{\rm SED}$}}
\newcommand{\chicomb}{\mbox{$X^2_{\rm SED+IMPRO}$}}
\newcommand{\mcore}{\mbox{$M_c$}}
\newcommand{\mstar}{\mbox{$m_*$}}
\newcommand{\sigmacl}{\mbox{$\Sigma_{\rm cl}$}}
\newcommand{\view}{\mbox{$\theta_{\rm view}$}}
\newcommand{\rcore}{\mbox{$R_{\rm c}$}}
\newcommand{\menv}{\mbox{$M_{\rm env}$}}

\newcommand{\methanol}{\mbox{CH$_{3}$OH}}

\newcommand{\methylcyanide}{\mbox{CH$_{3}$CN}}

\newcommand{\hcop}{\mbox{HCO$^{+}$}}

\shorttitle{}
\shortauthors{Yang et al.}

\begin{document}

\title{Image Profile (IMPRO) Fitting of Massive Protostars. I.\\Method Development and Test Cases of Cepheus A and G35.20-0.74N}

\author[0000-0001-8227-2816]{Yao-Lun Yang}
\affiliation{Star and Planet Formation Laboratory, RIKEN Pioneering Research Institute, Wako-shi, Saitama, 351-0198, Japan}
\email{yaolunyang.astro@gmail.com}

\author[0000-0002-3389-9142]{Jonathan C. Tan}
\affiliation{Department of Space, Earth \&\ Environment, Chalmers University of Technology, 412 93 Gothenburg, Sweden}
\affiliation{Department of Astronomy, University of Virginia, Charlottesville, Virginia 22904, USA}
\email{jonathan.tan@chalmers.se}

\author[0000-0003-4040-4934]{Rub{\'e}n Fedriani}
\affiliation{Instituto de Astrof\'isica de Andaluc\'ia, CSIC, Glorieta de la Astronom\'ia s/n, E-18008 Granada, Spain}
\email{fedriani@iaa.es}

\author[0000-0001-7511-0034]{Yichen Zhang}
\affiliation{Department of Astronomy, Shanghai Jiao Tong University, 800 Dongchuan Rd., Minhang, Shanghai 200240, People’s Republic of China}
\email{yczhang.astro@gmail.com}

\correspondingauthor{Yao-Lun Yang}
\email{yaolunyang.astro@gmail.com}

\begin{abstract}
Massive stars play a critical role in the evolution of galaxies, but their formation remains poorly understood. One challenge is accurate measurement of the physical properties of massive protostars, such as current stellar mass, envelope mass, outflow cavity properties, and system orientation. Spectral energy distribution (SED) fitting is widely-used to test models against observations. The far-infrared SED traces cold dust in envelopes, while the near- and mid-infrared (MIR) probes emission from outflow cavities and/or the inner envelope.  However, SED fitting has degeneracy limiting its ability to yield accurate measurements of protostellar properties.
Here, we develop image profile (IMPRO) fitting as a method to improve the characterization of protostars. We utilize brightness distributions from multi-wavelength MIR images of massive protostars taken by SOFIA/FORCAST  
as part of the SOFIA Massive Star Formation (SOMA) survey to constrain protostellar properties via comparison to a grid of radiative transfer models.
We develop a fitting pipeline to extract information along the outflow axis, which is then combined with the SED fitting to yield improved constraints on protostellar properties.
We apply the IMPRO fitting method on the nearby massive protostar Cepheus A, finding that its properties become more tightly constrained compared to SED fitting, especially in the inclination of the source. However, for the more distant G35.20-0.74N, we find that the spatial resolution of SOFIA/FORCAST limits the utility of this combined fitting pipeline. However, higher resolution MIR observations, e.g., with JWST, are expected to greatly expand the applicability of this fitting technique to protostars across the Galaxy.
\end{abstract}

\section{Introduction}

Massive stars (i.e., with mass $m_*>8 \msun$) shape galaxy evolution through intense radiation, powerful winds, supernovae, and nucleosynthetic enrichment \citep[e.g.,][]{2007prpl.conf..165B,2014prpl.conf..149T}. Constraining their formation processes is thus key to understanding their galactic impact. Their radiative and mechanical feedback also affects nearby molecular gas and the birth of lower-mass stars. Several formation theories have been proposed, including the Turbulent Core Accretion (TCA) model \citep{2003ApJ...585..850M} and the Competitive Accretion model \citep{2001MNRAS.323..785B}, which predict different observationally testable signatures. The turbulent core model expects a massive prestellar core supported by turbulence and/or magnetic fields that monolithically collapses under gravity, producing a structure similar to low-mass protostellar cores, i.e., a central disk with quasi-symmetric bipolar outflows. Competitive accretion instead predicts formation within a protocluster, where the global potential controls accretion onto the massive protostar, yielding more chaotic, disordered structure on small scales. Testing these models requires improved observational constraints on the fundamental properties of massive protostars.

Direct observations of massive protostars have been challenging due to their embedded nature.  While the extinction is much less at sub-mm and radio wavelengths, continuum observations in this regime mainly probe cold dust in the outer envelope or the midplane of the disk \citep{2016AARv..24....6B,2023ASPC..534..501M}. To better constrain protostellar properties, especially those related to luminosity, requires a complete multi-wavelength approach that utilizes the radiative transfer of MIR emission through the protostellar core.

\citet{2018ApJ...853...18Z} (hereafter ZT18) present a grid of radiative transfer simulations based on the TCA model, including a protostellar envelope, an accretion disk, a disk wind powered bipolar outflow, and a central protostar \citep{2011ApJ...733...55Z,2013ApJ...766...86Z,2014ApJ...788..166Z}.  The ZT18 model grid is spanned by five parameters: initial core mass ($M_c$); the mean mass surface density of the surrounding clump ($\Sigma_\text{cl}$); the current protostellar mass ($m_*$); the viewing angle, defined as the angle between line of sight to the near-facing outflow axis ($\theta_\text{view}$); and the equivalent V-band magnitude of foreground extinction ($A_V$). The first three parameters describe the physical models along protostellar evolutionary tracks, i.e., a separate track following an increasing $m_*$ for various combinations of $M_c$ and $\Sigma_{\rm cl}$ that describe initial core properties. The last two parameters aim to capture observational effects due to source orientation and foreground extinction. Use of the ZT18 grid has so far been limited to global SEDs, which are then fitted to observational data \citep[e.g.,][]{2017ApJ...843...33D,2019ApJ...874...16L,2020ApJ...904...75L}.  In addition, \citet{2023ApJ...942....7F} implemented ZT18 SED fitting via a python package, \texttt{sedcreator}\footnote{https://sedcreator.readthedocs.io/en/latest/}. \citet{2025ApJ...986...15T} have developed methods to identify and fit SEDs of multiple sources in crowded regions.

However, the observed SED may not uniquely constrain model parameters. For example, ``good'' models returned by SED fitting, i.e., with acceptable reduced $\chi^2$ values, can span large ranges of model grid parameters \citep[see, e.g.,][]{2018ApJ...853...18Z,2023ApJ...942....7F}. Spatial brightness distributions provide independent constraints that can break such degeneracies and reduce uncertainties in derived protostellar properties. \citet{2011ApJ...733...55Z,2013ApJ...766...86Z,2014ApJ...788..166Z} demonstrated distinct brightness profiles along the outflow direction at different viewing angles, suggesting the constraining power of radial brightness profiles for the ZT18 grid parameters. \citet{2013ApJ...767...58Z} presented an example of fitting brightness profiles along a massive protostellar outflow. Using a different approach, \citet{2019AA...625A..44F} showed that multi-scale observations, with the highest resolution probing down to the disk scale, can more robustly constrain models to reproduce both the SED and brightness distributions at different scales. A larger survey using this method further explored disk evolution in massive protostars \citep{2021AA...648A..62F,2021ApJ...920...48F}.

Here, we present a modeling method for massive protostars, fitting both the SED and the radial brightness profile along the outflow axis with the ZT18 grid. Section\,\ref{sec:method} describes the fitting method and how we combine the SED and the radial profiles. Section\,\ref{sec:results} presents two case studies of massive protostars, Cepheus A and G35.20$-$0.74N, which we fit by this method. We also discuss the performance of our new fitting method compared with SED-only fitting. Finally, Section\,\ref{sec:conclusions} summarizes the conclusions of our study.

\section{Method}

Uncertainty in viewing angle is one of the major sources of degeneracy in fitting SED models to observed protostars.  For low-mass protostars, where the disk can be resolved, radiative transfer modeling of the resolved disk emission together with the SED modeling has shown success in constraining the physical properties of the cores and disks, especially the disk mass \citep{2022ApJ...929...76S}.  While the disks of massive protostars are often unresolved, limiting a similar application, the outflows provide another diagnostic to constrain the viewing angle.  As a result of low-density outflow cavities, the SED of a dense envelope viewed from the outflow direction can appear similar to the SED of a less dense envelope viewed closer to the midplane of the disk (edge-on view).  This issue leads to degeneracy between the viewing angle and the core properties in SED models.  While ZT18 presents a self-consistent SED model grid of massive protostars, the degeneracy due to the viewing angle inevitably hinders an accurate characterization of a protostar.  Outflow morphology provides a way to break this degeneracy.  A protostellar envelope with an inclined outflow cavity would have uneven brightness between the two outflow lobes.  The asymmetric brightness becomes more obvious at shorter wavelengths, where extinction is notably higher toward the red-shifted, far-facing outflow.  Thus, the asymmetry in the image may provide unique constraints on the inclination and help to break degeneracies in SED model fitting.

\subsection{IMPRO}\label{sec:method}

\begin{figure*}[htbp!]
  \centering
  \includegraphics[width=\textwidth]{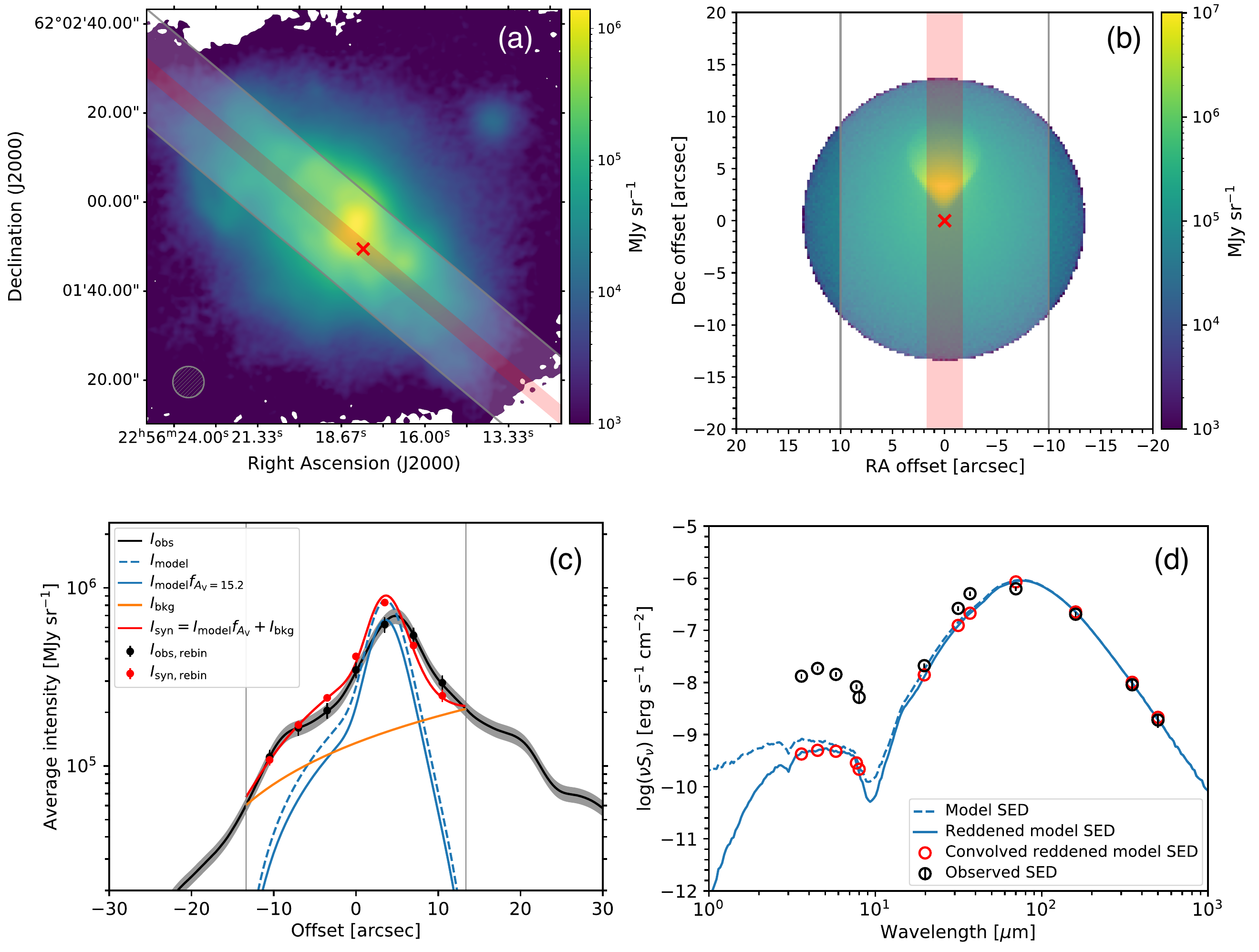}
  \caption{Illustrations of steps in the combined IMPRO + SED fitting pipeline, using Cep A as an example.  (a) The 37 \micron\ SOFIA/FORCAST image (see Section\,\ref{sec:obs}), as an example, overlaid with two strips of 3.5\arcsec\ and 20\arcsec\ width (red and white) along the outflow direction (PA=50\arcdeg).  The beam size is shown at the bottom left, and the protostar is marked with a red ``x'' (defined from radio continuum emission).  (b) The 37 \micron\ synthetic image overlaid with the same two strips of 3.5\arcsec\ and 20\arcsec\ width.  The source is centered at the red ``x'' and the outflow axis is aligned N-S.  (c) The extracted 1D brightness profiles along the strips.  Each profile can be related using Equation\,\ref{eq:fitting}.  The \chisq\ value is derived by comparing the rebinned observed profile (black circles) against the rebinned synthetic profile (red circles).  Panels (c) and (d) show the best-fitting model considering both the 1D profile and the SED.  The \av\ adopted in this example is 15.2~mag.  (d) The observed SED is compared with the synthetic SED of the best-fitting model.  The black and red circles show the fluxes at the wavelengths where photometry is available.  The synthetic SED (red circles) is derived by reddening the SED convolved with photometric filters.  The blue solid and dashed lines show the synthetic SED with and without extinction.}
  
  \label{fig:workflow}
\end{figure*}

We develop an image profile (IMPRO) fitting method to better constrain the properties of protostellar cores by fitting for the brightness profile along the outflow together with the SED modeling developed by ZT18.  Figure\,\ref{fig:workflow} illustrates steps in this method.  The IMPRO fitting uses infrared images to constrain the brightness morphology due to the outflow and its inclination. To limit the number of degrees of freedom, we extract the brightness profile from a strip along the outflow axis. The strip has a default width of 20\arcsec\ (see the discussion in Section\,\ref{sec:cepA_fitting}).. Averaging the brightness along the width of the strip, we can extract a 1D brightness profile.  To constrain the protostellar model, we perform the same extraction at all models in the ZT18 grid.  In ZT18, only the synthetic SEDs are presented.  However, the model images are also produced as part of the radiative transfer calculations.  Example model images were presented in \citet{2014ApJ...788..166Z}.  The model grid assumes a distance of 1~kpc, so we first calculate the distance-corrected angular size of the model image and then extract the 1D profile with a strip of the same width. For partially included pixels, the calculations of the average brightness take the fractional brightness according to the included area.  Figure\,\ref{fig:workflow}c shows the extracted 1D profile from an observation and a protostellar model.. The details of the observational data used in this study is described in Section\,\ref{sec:obs}.  The full fitting pipeline includes images at multiple wavelengths (see Section\,\ref{sec:cepA_fitting}).

Being located at large distances of $\gtrsim 1$ kpc, massive protostars often have significant foreground extinction, which needs to be accounted for when comparing observed systems with theoretical models.  The ZT18 SED model grid considers the extinction (\av) separately by minimizing the $\chi^2_\text{SED}$ in each model SED with \av\ ranging from 0 to 1000 \citep{2023ApJ...942....7F}, but the model images do not include the effect of \av.  Furthermore, the model only includes the protostellar core, whereas the core often resides within a clump, which emits significant ``background'' emission.  Thus, to realistically compare the modeled 1D profiles ($I_{\rm model}$) with the observed 1D profile ($I_{\rm obs}$), we need to consider the effects of such foreground extinction and background emission.

For each model, we consider the emission in observational data from outside of the model core radius (Equation\,\ref{eq:rc}; see also ZT18) as the background emission.  The background emission is determined at the core radius ($\pm \rcore$ in the 1D profile) and linearly interpolated along the strip. In the TCA model, the core radius is given by 
\begin{equation}
  \frac{R_{\rm c}}{\text{pc}} = 5.7\ee{-2} \left(\frac{M_c}{60\,M_\odot}\right) \left(\frac{\Sigma_{\rm cl}}{\text{g cm}^{-2}}\right)^{-0.5}.
  \label{eq:rc}
\end{equation}
The estimated background emission from observations is the background emission with foreground extinction.  To compare the model against the observations, we apply the foreground extinction to the modeled intensity ($I_{\rm model}$) and include the estimated background emission ($I_{\rm bkg}$), resulting in a synthetic 1D brightness profile ($I_{\rm syn}$).  The relations between these variables are given by 
\begin{align}
  I_{\rm syn} & = I_{\rm model}f_{A_{\rm V}} + I_{\rm bkg}; \nonumber \\
  f_{\rm A_{\rm V}} & = 10^{-0.4A_V(\kappa_{\lambda}/\kappa_V)}, 
  \label{eq:fitting}
\end{align}
where the $\kappa_{V}$ is the dust opacity at V-band (0.55 \micron).
The extinction relation follows the prescription in ZT18, which uses the extinction law in \citet{1994ApJ...422..164K}.  

We consider \av\ as an independent parameter and sample it in a natural logarithmic scale from 1 to 1000~mag with 100 points, in addition to \av=0.  Thus, there are five parameters in the protostellar models, core mass (\mcore), mass surface density (\sigmacl), protostellar mass (\mstar), viewing angle of line of sight to outflow axis (\view), and foreground extinction (\av).  ZT18 sampled the first three parameters with 432 models to construct the grid of SED models.  The viewing angle was sampled linearly in cos\view\ with 20 points from 0\arcdeg\ to 90\arcdeg.  As demonstrated by \citet{2013ApJ...766...86Z}, the brightness profile along the outflow is sensitive to \view, showing asymmetry between the two sides of the outflow when the source is viewed at an angle (i.e., not edge-on nor pole-on).  We further show the effect of \view\ in the synthetic 1D brightness profile in Appendix\,\ref{sec:syn_view}.  In total, the model grid has 8640 models without considering \av.  To find the best-fitting models, we derive the $X^2$ of each model by comparing it with the 1D profile extracted from observations.  Because the uncertainty ($\sigma$) may be systematic, for example, due to calibration, which may not follow Gaussian distribution, we use $X^2$ to represent of difference between model and data, specifically highlighting a $\chi^2$ distribution only under restricted conditions \citep{chernoff_use_1954,watson_chi-square_1958,cressie_pearsons_1989}. To avoid over-sampling the spatial resolution, we then resampled the 1D profiles of $I_{\rm obs}$ and $I_{\rm syn}$ onto a coarse grid with each element separated by the spatial resolution of the observations, resulting in $I_{\rm obs,bin}$ and $I_{\rm syn,bin}$.  The rebinned width depends on the resolution of photometric images, which is further discussed in Section\,\ref{sec:obs}.  Furthermore, we only consider the brightness within the model $R_{c}$.  For each model, we calculate the reduced \chioned\ as
\begin{align}
  X_{\rm IMPRO}^2 &= \frac{1}{n_{\rm bin}-1}\sum\left(\frac{(I_{\rm obs, bin}-I_{\rm syn, bin})^2}{\sigma^2}\right)\text{, where} \nonumber \\
  \sigma &= \sqrt{(0.1I_{\rm obs, bin})^2+ (0.1I_{\rm bkg, bin})^2},
  \label{eq:chi_impro}
\end{align}
assuming a 10\%\ systematic calibration uncertainty on $I_{\rm obs}$ and $I_{\rm bkg}$, consistent with the assumption adopted in the SOMA papers \citep[e.g.,][]{2017ApJ...843...33D}, where the SED fitting approach was developed.  The $n_{\rm bin}$ is the number of resampled data points along the 1D profile within $R_{\rm c}$.  The rebinned data points, $I_{\rm obs, bin}$, $I_{\rm bkg, bin}$, and $I_{\rm syn, bin}$, are considered as separate measurements in assessing the \chioned\ when multiple images are considered.  In Equation\,\ref{eq:chi_impro}, the summation is performed over all rebinned data points across all wavelengths.  Thus, the 1D profiles at the shorter wavelengths are weighted slightly higher than that at the longer wavelengths due to their finer sampling over the same offset.  We choose this weighting approach to increase the sensitivity to \view\ and \av, both of which are better probed by the images at shorter wavelengths.  

While the bin size is chosen to match the spatial resolution, the number of rebinned data points is adaptively determined depending on the core size in the model (i.e., the number of rebinned data is not fixed), the degree-of-freedom, represented by $n_\text{bin}$ in Equation\,\ref{eq:chi_impro}, is ill-defined; thus, the $X^2$ statistics and probabilities cannot be unambiguously assigned \citep{tibshirani_degrees_2015}.  The effect of this adaptive rebinned process on the fitting result is discussed in Section\,\ref{sec:limitation} and will be further tested and characterized in follow-up studies.  We note the impact on our fitted values in Table\,\ref{tbl:best_params} and the significant figures of the fitted values are conservatively estimated.  This pipeline is designed to model outflow-tracing emission, which requires sufficient S/N.  Applications on faint sources may not yield accurate statistical estimates.

In addition to \chioned, we use the modeling pipeline in \citet{2023ApJ...942....7F} to derive the \chised\ from the SED using the full set of parameters (Figure\,\ref{fig:workflow}d).  Here we also use \chised\ instead of $\chi^2_\text{SED}$ to indicate that this statistic may not strictly follow the $\chi^2$ statistics.  The original pipeline in ZT18 steps \av\ from 0 to 100~mag for each combination of \mcore, \sigmacl, \mstar, \view\ and only reports the best-fitting model.  \citet{2023ApJ...942....7F} revised the SED modeling pipeline, called \texttt{sedcreator}, to (1) find the best-fitting \av\ from 0 to 1000~mag with a minimization algorithm for combination of \mcore, \sigmacl, \mstar, \view; and (2) convolve the synthetic SED with photometric filters before applying the effect of \av, while the original ZT18 pipeline performs these two steps in the opposite order, resulting in small but noticeable differences \citep{2023ApJ...942....7F}.  While \texttt{sedcreator} allows us to treat \av\ as an independent parameter, consistent with the treatment in the fitting of the 1D radial profiles, these two pipelines differ in their sampling of the same range of \av.  The \texttt{sedcreator} algorithm finds an \av\ from 0 to 1000~mag that best fits the SED, whereas the IMPRO pipeline uses a logarithmic grid of 100 points from 1 to 1000~mag, as well as considering \av=0.  Thus, we modified \texttt{sedcreator} to perform a grid search using the same \av\ grid of the IMPRO pipeline.  Comparing with the default fitting results of the \texttt{sedcreator}, the \chised\ of the best-fitting model for each combination of \mcore, \sigmacl, \mstar, \view\ only differs $\lesssim 6\%$ due to the slight mismatch of the \av\ grid.  In the SED fitting, we treat the mid-infrared fluxes at $<$10 \micron\ (mostly the Spitzer/IRAC photometry) as upper limit because PAH emission, which is not considered in the modeling, can contaminate the photometric fluxes.  This treatment is consistent with that adopted in the SED modeling \citep{2023ApJ...942....7F}.

To assess the goodness of fit, we weight the 1D profile and the SED equally by averaging the \chioned\ and \chised\ to have a combined \chicomb.  This simple approach considers the IMPRO and SED fittings as two independent estimates of the model parameters.  As we discuss in Section\,\ref{sec:comb_fitting}, the two fitting approaches are, indeed, sensitive to different model parameters.  Given the test cases presented in this study, this simple treatment already provides reasonable estimates of the protostellar cores.  Thus, we treat the \chisq\ in the IMPRO and SED fitting equally in estimating the goodness of the combined fit.  Following a similar strategy as that in the SED modeling \citep{2025ApJ...986...15T}, we define ``good'' models as those with their \chicomb\ no greater than the minimum $\chicomb+2$.  The SED fitting of \citet{2025ApJ...986...15T}  uses the same ZT18 model grid and a similar definition of ``good'' models, which allow us to better compare these previous SED-only results with the fitting presented in this study.  The IMPRO and IMPRO+SED fitting pipelines are also benchmarked against synthetic observations using the ZT18 model image, showing that consistent model parameters can be retrieved using the pipelines (Appendix\,\ref{sec:benchmark}).

\section{Results and Analysis}\label{sec:results}

We applied the IMPRO fitting pipeline to two well-studied sources in the SOMA survey, Cepheus A (hereafter Cep A) and G35.20$-$0.74N (hereafter G35.2N) \citep{2017ApJ...843...33D}.  Cep A is one of the closest massive protostars ($d\sim700\:$pc;  \citealt{2009ApJ...693..406M}) and has well-characterized outflows, so that the asymmetric brightness distribution due to the outflows is well-resolved by SOFIA-FORCAST.  G35.2N, on the other hand, is further away at 2.2 kpc \citep{2009ApJ...693..419Z}. \citet{2013ApJ...767...58Z} fitted the G35 1D brightness profile along the outflow direction using a model based on an earlier version of the ZT18 model grid but with extended outflow and clump environment structures.

As mentioned in Section\,\ref{sec:method}, we ran the fitting pipelines (IMPRO and SED) and selected the ``good'' models to constrain the model parameters, inferring the physical properties of the massive protostellar cores.  The interpretation of the fitted parameters has two challenges, the non-linear sampling of \mcore, \mstar, and \sigmacl, and the large range of values sampled in some parameters (e.g., \av).  To address this first point, the comparison is performed by calculating the geometric mean  (averaged in logarithmic scale) for \mcore, \mstar, and \sigmacl\ and the arithmetic mean for \view\ and \av, in the good models.  This method follows the SED fitting procedures developed in the series of papers using the same model grid \citep{2017ApJ...843...33D,2019ApJ...874...16L,2020ApJ...904...75L,2023ApJ...942....7F,2025ApJ...986...15T}.  The difference in the averaging methods comes from the sampling of these model parameters in the ZT18 grid, where the \mcore, \mstar, and \sigmacl\ are sampled non-linearly, while \view\ and \av\ are sampled linearly.  The second point is addressed partially by taking the geometric means.  To fully capture the potentially large ranges of parameters in the good models, we also use the minimum and maximum values in the parameters to represent the possible range of the fitted parameters.

\subsection{Observational Data}\label{sec:obs}
\begin{figure*}[htbp!]
    \centering
    \includegraphics[width=\textwidth]{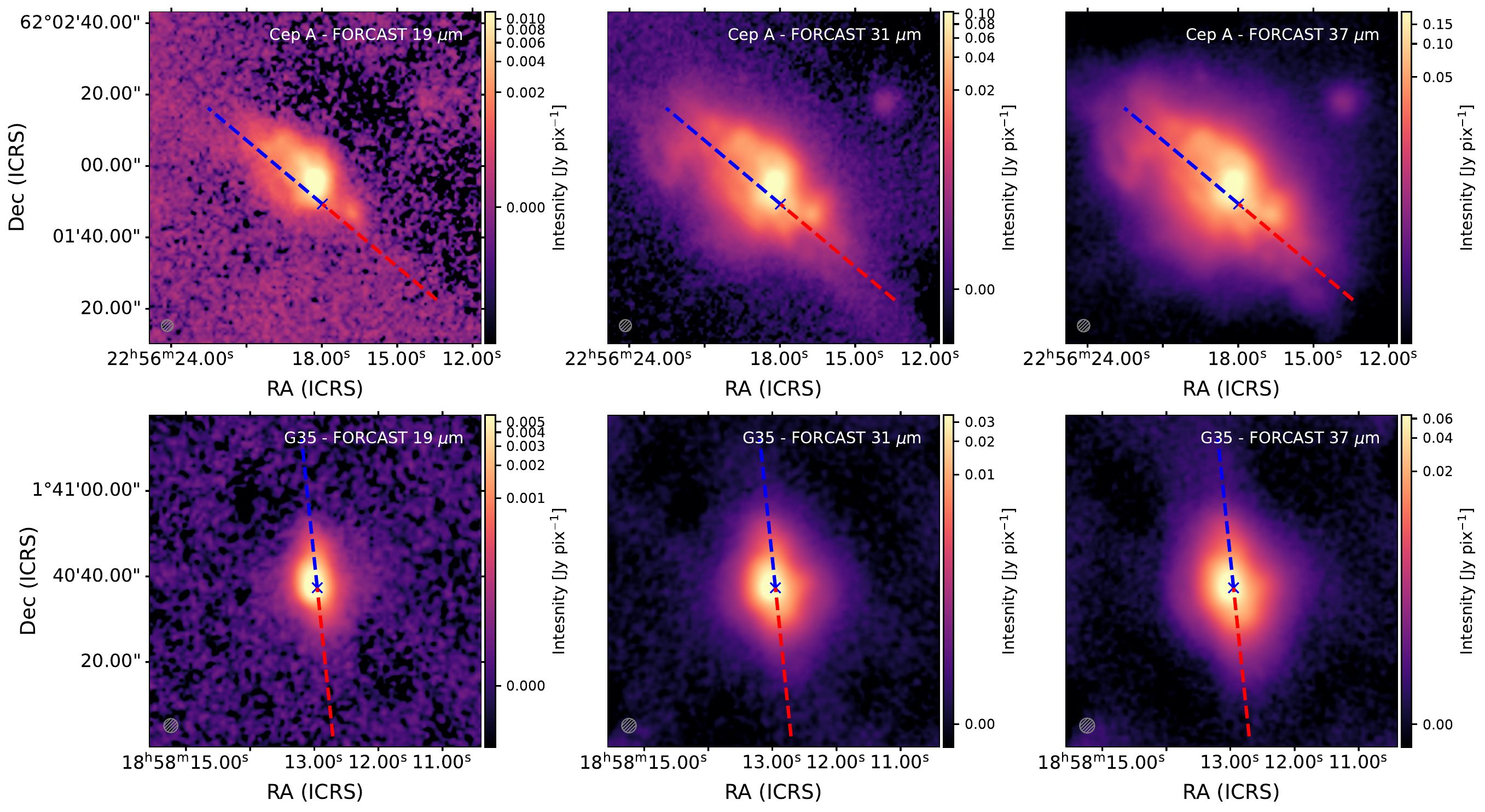}
    \caption{SOFIA FORCAST images of Cep A (top) and G35.2N (bottom) at 19.7, 31.5, and 37.1 \micron\ (from left to right).  The blue and red dashed lines indicate the blue- and red-shifted outflow directions.}
    \label{fig:forcast}
\end{figure*}

We first introduce the observational data used in this study, providing the context to justify the choice of fitting parameters the following sections.  We use both SED and infrared images as the data for fitting.  The IMPRO fitting uses the SOFIA/FORCAST infrared images taken by the SOMA survey \citep{2017ApJ...843...33D}.  While the SOMA survey contains images from various filters ranging from 7.7 to 37.1 \micron.   We exclude the shorter wavelength filters, such as the 7.7 and 11 \micron\ filter from the SOMA survey because of the possible contamination of the polycyclic aromatic hydrocarbon (PAH) emission.  The resulting three filters used in this study are the 19 \micron\ ($\lambda_\text{eff}=19.7 \micron$, $\Delta\lambda=5.5 \micron$), 31 \micron\ ($\lambda_\text{eff}=31.5 \micron$, $\Delta\lambda=5.7 \micron$), and 37 \micron\ ($\lambda_\text{eff}=37.1 \micron$, $\Delta\lambda=3.3 \micron$).  Figure\,\ref{fig:forcast} shows the SOFIA FORCAST images from the SOMA survey used by the modeling presented in this study \citep{2017ApJ...843...33D}.  In this study, we refer these images by their effective wavelengths at 19.7, 31.5, and 37.1 \micron.  The beam sizes at these images are 3\farcs{20}, 2\farcs{85}, and 2\farcs{54} for 37.1, 31.5, and 19.7 \micron\ images, respectively\footnote{https://irsa.ipac.caltech.edu/data/SOFIA/docs/instruments/\\handbooks/FORCAST\_Handbook\_for\_Archive\_Users\_Ver1.0.pdf}.  The adopted rebinned sizes for these three filters are chosen to be 3\farcs{5}, 3\arcsec, and 2\farcs{5} for 37.1, 31.5, and 19.7 \micron\ images, respectively.  The SEDs are taken from \citet{2023ApJ...942....7F}, where they collected public archival photometric data from Spitzer/IRAC \citep{2004ApJS..154...10F,2004ApJS..154....1W}, SOFIA/FORCAST \citep{2017ApJ...843...33D}, Herschel/PACS and SPIRE \citep{2010AA...518L...3G,2010AA...518L...2P}.  They extracted the SED values using an algorithmically optimized aperture (see Section 3.1.1 in \citealt{2023ApJ...942....7F}).  The apertures for CepA and G35.2N are 72\farcs{5} and 37\farcs{5}, respectively.

\subsection{Cepheus A}
\label{sec:cepA_all}

Located at 700 pc, Cep A is the second closest massive star-forming region (with Orion being the closest at 400 pc). Thus, photometric surveys, such as SOMA, have a better physical resolution ($\sim$2000 au) at Cep A to probe the asymmetric warm dust continuum due to outflows.  In the development of the fitting pipeline, we thus used Cep A to optimize the choice of the strip width and the combination of photometric bands, which are discussed in the following sections.

\subsubsection{The Outflows and Protostars in Cep A}
Cep A has a main bipolar outflow axis oriented in the east-west direction, with a full extent of $\sim$1 pc, first detected by \citet{1980ApJ...240L.149R}.  While the large-scale outflows orient in the E-W direction, the compact emission near Cep A traces outflows in the NE-SW direction.  Together with the bow shocks traced by H$_2$, \citet{2009ApJ...692..943C} infer a pulsed, precessing jet in Cep A.  At high velocity, $20 < |v-v_{\rm source}| < 70$ \kms, the CO emission shows a compact bipolar outflow morphology with a position angle (PA) of about 40\arcdeg--60\arcdeg\ with a length of $\sim$0.4 pc.  The compact outflow also appears in \hcop\ \jj{1}{0} emission, as well as in thermal radio emission.  The \hcop\ outflows have velocities up to $\sim$50 \kms, extending $\sim$1\arcmin\ ($\sim$0.2 pc) with a PA of 55\arcdeg--60\arcdeg\ \citep{1999ApJ...514..287G}.  The MIR emission focused in this study traces the compact NE-SW outflows.  We adopt an outflow PA of 50$^\circ$.  The choice of PA has little impact on the physical parameters of the core (\mcore, \sigmacl, and \mstar) but has some effect on \view\ and \av\ (Appendix\,\ref{sec:pa_test}).

\citet{1984ApJ...276..204H} identified a cluster of radio sources at the center of the bipolar outflows, where HW2 is the strongest source with a flux density of 15.8$\pm$0.3 mJy at 14.9 GHz \citep{1996ApJ...459..193G}.  High resolution and multi-frequency observations suggest HW2 as a biconical thermal jet with a proper motion of $\sim$500 \kms\ \citep{2006ApJ...638..878C}.  Among the radio sources, only HW2 and HW3c exhibit strong continuum at 870 \micron, suggesting that HW2 is the most luminous protostar, followed by HW3c \citep{2007ApJ...660L.133B}.  The total luminosity of Cep A has been previously estimated to be 2.5\ee{4} \lsun\ \citep{1981ApJ...244..115E}. The SOMA SED fitting result for Cep A derives an intrinsic bolometric luminosity of $L_{\rm bol}=4.9^{+2.8}_{-1.8} \times 10^4$ \lsun\ and an isotropic bolometric luminosity (based on the observed bolometric flux) of $L_{\rm bol, iso}=2.0^{+0.5}_{-0.4} \times 10^4$ \lsun, with these uncertainties reflecting the dispersion in the ``good'' SED model fits \citep{2025ApJ...986...15T}.  In the SOMA SED fitting, the $L_{\rm bol}$ is the total luminosity radiated by the star and the accretion layer, while the $L_{\rm bol, iso}$ is the luminosity emitted by the protostellar core, considering the effect of its geometric structure, such as outflow cavities.  The foreground extinction is not considered for $L_{\rm bol, iso}$.

Many studies consider HW2 as the driving source of the complex outflows in Cep A. In our analysis, we will use the radio position of this source to define the location of the protostar \citep[see][]{2017ApJ...843...33D,2019ApJ...873...20R}. Positive values of offset are defined to be in the direction of the near-facing, blue-shifted outflow axis.  Observations of the 2.12 \micron\ H$_2$ trace several Herbig-Haro (HH) objects along the NE-SW outflows.  To the west of HW2, HH 168 (also known as GGD 37; \citealt{1978ApJ...224L.137G,1999ASIC..540..267R}) is the brightest HH object in the complex \citep{1986AJ.....92.1155H,2000AJ....120.1436H,2011ApJ...726L...1G}.  The precessing jet inferred by \citet{2009ApJ...692..943C} also considers HW2 as the driving source, which results in most of the HH objects to the east and northeast, while the HW3c may be the driving source of HH 168. 

\citet{2017AA...603A..94S} derived an outflow inclination (\view) of 26\arcdeg\ by modeling a planer motion from observations of \methanol\ maser.  \citet{2005Natur.437..109P} derived a disk inclination of 62\arcdeg\ from fitting a Keplerian motion to the \methylcyanide\ emission, suggesting an outflow inclination (\view) of 28\arcdeg\ if the outflow is perpendicular to the disk.  They also derived a binding mass of 19$\pm$5 \msun\ from the observed line width.  However, \citet{2007ApJ...660L.133B} discovered distinct chemical differentiation along the presumed disk major axis, suggesting the presence of multiple hot cores at a separation of $\sim$1\arcsec\ instead of a disk.  Interestingly, \citet{2013arXiv1305.4084Z} proposed that the precession at large scales may be due to the dynamical interactions between companions.  The dust emission remains elongated at 0\farcs{6}, supporting the existence of a disk, while multiple hot cores could exist in the vicinity of the dominant protostar \citep{2007ApJ...666L..37T}.  At $\sim$30 mas ($\sim$20 au) resolution, \citet{2021ApJ...914L...1C} found a highly collimated bipolar feature at 40 GHz that can be reproduced with a collimated jet along with a wide-angle wind, which was also suggested by \citet{2011MNRAS.410..627T}.  The abundant observational constraints on both the central protostar and the small-scale ($<1\arcmin$) outflow make Cep A an ideal target to test the fitting of the radial brightness profiles.

\subsubsection{IMPRO Fitting the Outflow Axis Intensity Profiles}
\label{sec:cepA_fitting}

\begin{figure}[htbp!]
  \centering
  \includegraphics[width=0.47\textwidth]{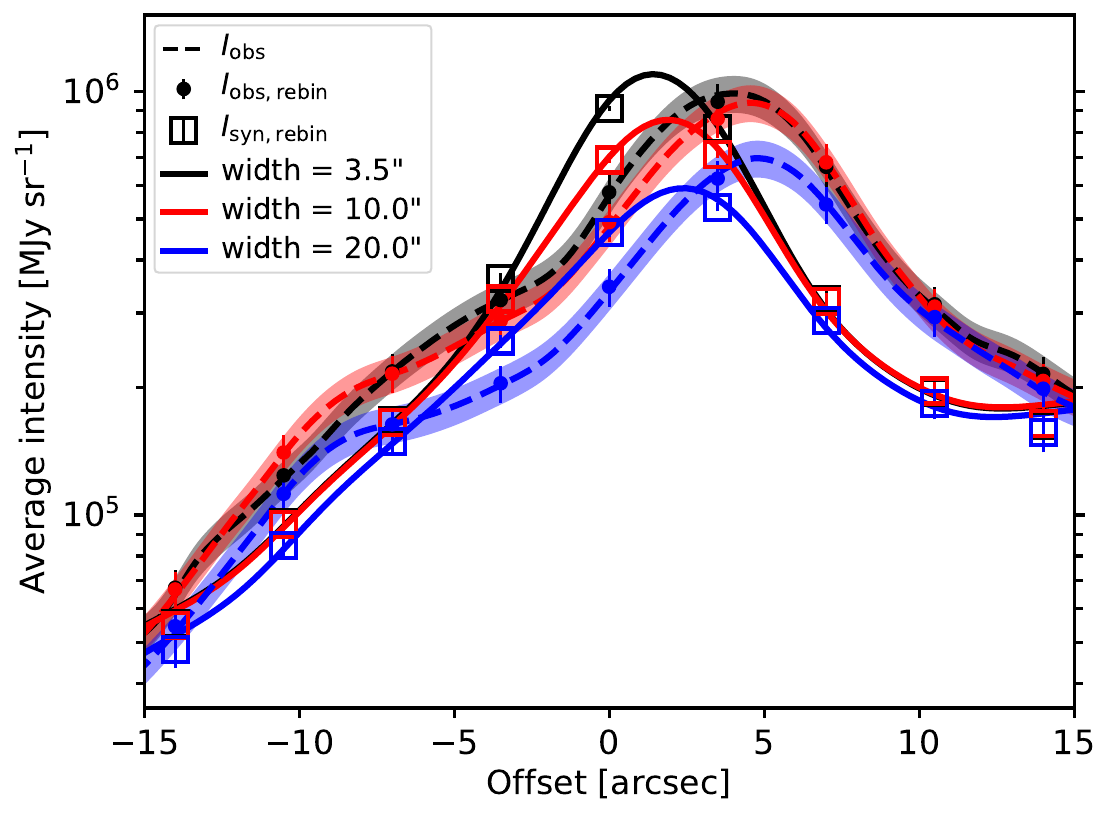}
  \caption{The Cep A 1D observed (dashed) and synthetic (solid) 37.1~\micron\ image intensity profiles derived with strip widths of 3\farcs{5} (black), 10\arcsec\ (red), 20\arcsec\ (blue).  The physical model has \zt{160}{3.16}{64}{74} and is the same for all three sets of profiles.  The rebinned observed and synthetic intensities are shown in filled circles and open squares, respectively.}
  \label{fig:width_1dcomp}
\end{figure}

\begin{figure*}[htbp!]
  \centering
  \includegraphics[height=1.34in]{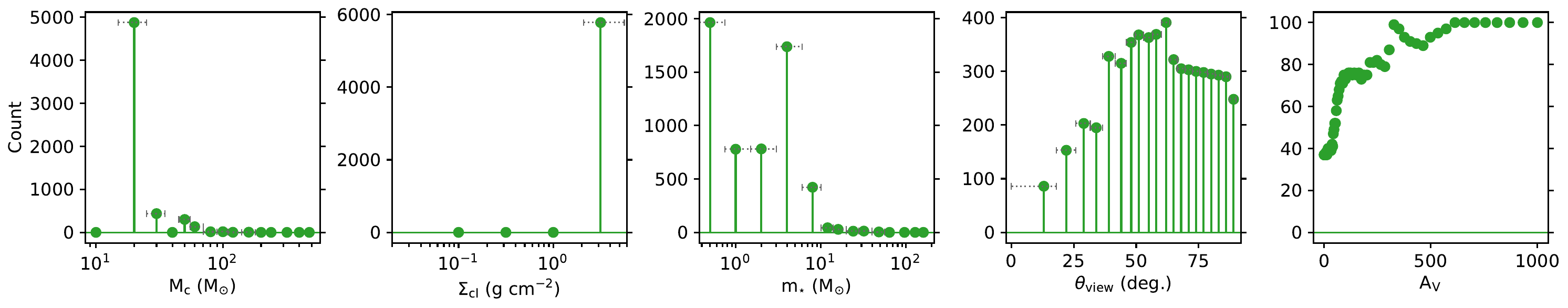}
  \includegraphics[height=1.34in]{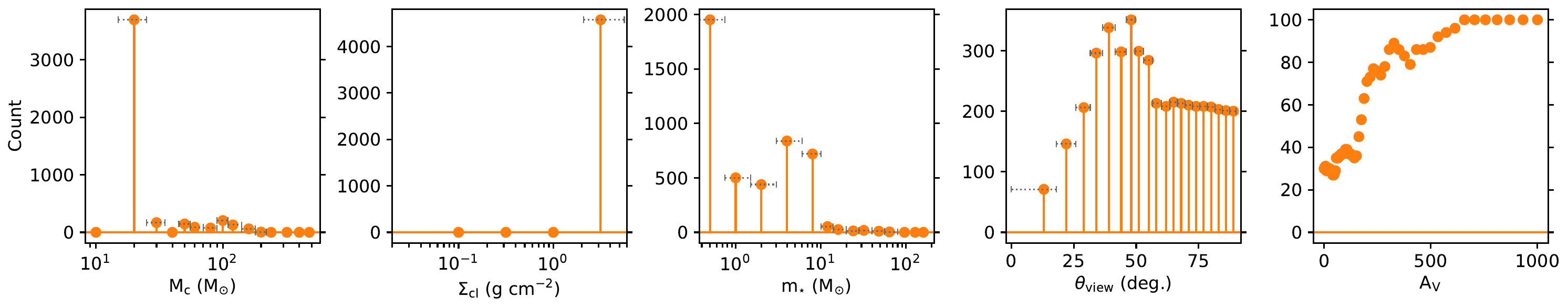}
  \includegraphics[height=1.34in]{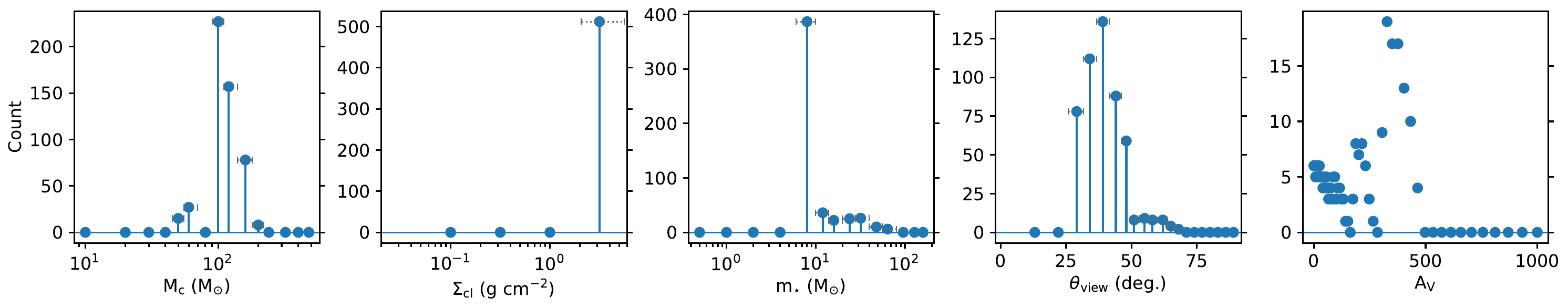}
  \caption{Distributions of the Cep A model parameters of the good 1D profile models ($X^2_\text{min, IMPRO} < X^2_\text{IMPRO} < X^2_\text{min, IMPRO}+2$).  The strip widths are 3.5\arcsec, 10\arcsec, and 20\arcsec\ from top to bottom.  The gray horizontal bars in the first four columns show the range of the parameter value that is represented by the particular radiative transfer model in the ZT18 model grid, i.e., extending to the logarithmic midpoints between adjacent sampled models in \mcore, \sigmacl, and \mstar\ and linear midpoints in cos\view.  These fittings include only the 37.1 \micron\ image.}
  \label{fig:width_comp}
\end{figure*}

\textbf{Strip width - } While Section\,\ref{sec:method} outlines the prescription of fitting the radial brightness profile, we need to first determine the fitting setup, such as the strip width and the choice of photometric bands, to have the best fitting performance.  A good choice of the strip width would robustly constrain the viewing angle (\view).  We tested three strip widths, 3\farcs{5}, which is similar to the size of the SOFIA beam at 37 \micron, 10\arcsec, and 20\arcsec.  The width of the latter two strips is larger than the substructures seen in the outflows (Figure\,\ref{fig:workflow}a).  We use the ``good'' models, defined in Section\,\ref{sec:method}, to examine the performance of the fitting with different strip widths.  This examination aims to determine the ideal strip width for fitting the 1D profiles, thus focusing only on the \chisq\ calculated from the extracted 1D profile (\chioned).  The difference of the peak positions between the observed and synthetic 1D profiles reduces when the strip width increases, suggesting a better fit (Figure\,\ref{fig:width_1dcomp}).  Figure\,\ref{fig:width_comp} shows the distributions of all five model parameters among the good models for three different strip widths.  The bin sizes reflect the intrinsic sampling in the ZT18 model grid.  The best-fitting model parameters span wide ranges of values in the fitting using a narrow strip.  For example, the \view\ values span the entire possible range (0\arcdeg--90\arcdeg) in the fittings with 3\farcs{5} and 10\arcsec\ wide strips, whereas the fitting with a 20\arcsec\ wide strip results in a narrow distribution of \view.  The best-fitting \view\ distributions of all three fittings peak at a similar value, suggesting that they produce consistent results; however, the fitting with a 20\arcsec\ wide strip provides more diagnostic constraints.  Thus, we define the strip width as 20\arcsec\ for the results shown in the following sections. More generally, the choice of strip width will depend on various source properties, especially distance, so it should be considered to be a user-defined parameter of the fitting procedure.

\textbf{Photometric bands - }The asymmetric brightness due to the orientation of outflow cavities would appear in the images at the wavelengths sensitive to warm dust emission and/or extinction.  Therefore, we can leverage the multi-band images acquired by the SOMA survey at 19.7, 31.5, and 37.1 \micron\ to explore the benefits of a combined fitting of the 1D profiles in multi-band images.  In the multi-band fitting pipeline, we first extracted the 1D observed and synthetic profiles in each band.  We rebinned the 1D profiles to 2\farcs{5}, 3\farcs{0}, and 3\farcs{5} for 19.7, 31.5, and 37.1 \micron, respectively, better matching with their corresponding resolutions.  Then, we calculated a reduced \chioned\ over three 1D profiles with equal weighting of each rebinned data point.  

\begin{figure*}
  \centering
  \includegraphics[width=\textwidth]{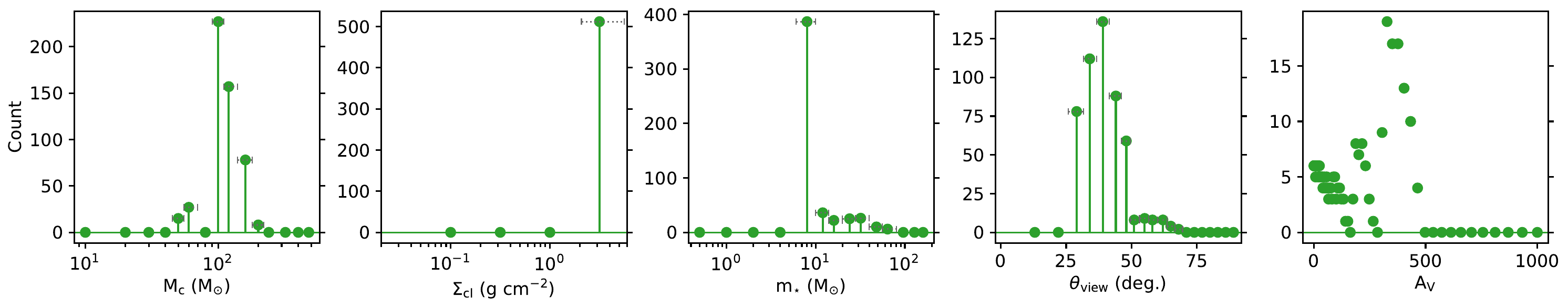}
  \includegraphics[width=\textwidth]{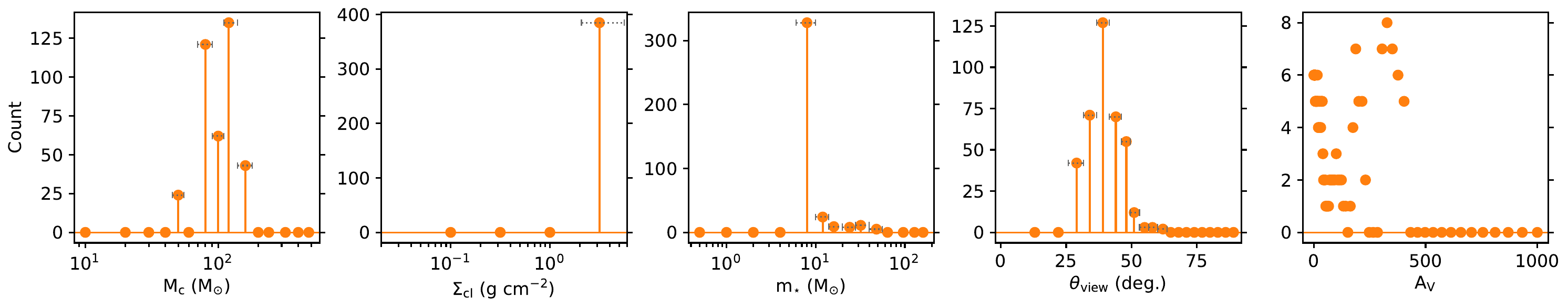}
  \includegraphics[width=\textwidth]{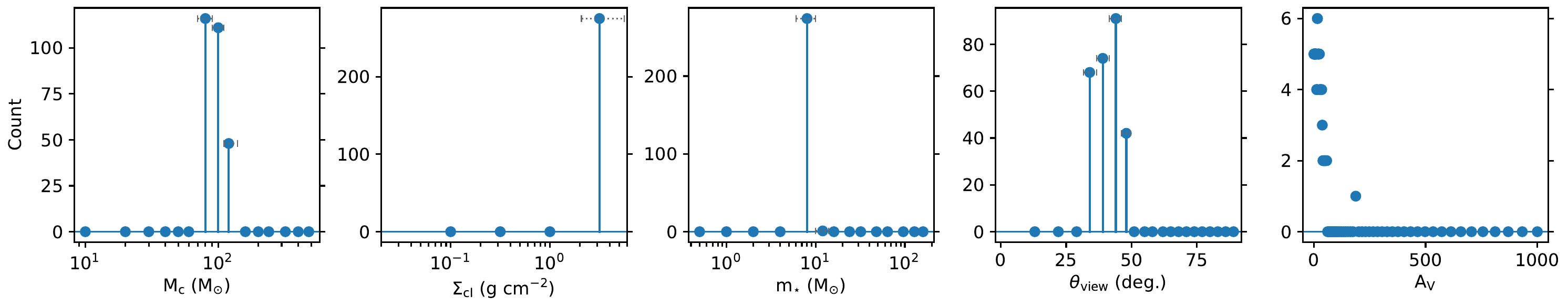}
  \caption{Distribution of the Cep A model parameters of the good 1D profile models ($X^2_\text{min, IMPRO} < X^2_\text{IMPRO} < X^2_\text{min, IMPRO}+2$) by fitting only 37.1~\micron, 37.1~\micron\ + 31.5~\micron, and 37.1 \micron\ + 31.5~\micron\ + 19.7~ \micron\ images (from top to bottom) using a 20\arcsec\ strip width.  Similar to Figure\,\ref{fig:width_comp}, the gray horizontal bars show the range of the parameter value that is represented by the particular radiative transfer model in the ZT18 model grid.}
  \label{fig:band_selction_comp}
\end{figure*}

Simultaneous fitting of the 1D profiles from three FORCAST bands, 19.7, 31.5, and 37.1 \micron, provides better diagnostic constraints on the protostellar model.   Figure\,\ref{fig:band_selction_comp} shows the model parameters of the good models fitted with three choices of band combinations: 37.1~\micron; 37.1~\micron + 31.5~\micron; and 37.1~\micron + 31.5~\micron + 19.7~\micron.  The fittings constrain the model parameters within a relatively narrow range of values regardless of the choice of band combination, except for \av.  The fittings with only 37.1 \micron\ and 37.1 \micron+31.5 \micron\ yield similar results, suggesting that the emission at these two bands comes from similar structures.  By including all three bands, the number of good models at higher \av\ reduces, and then the upper end of the \av\ range reduces.  While the \av\ in the good models has no dominant peak, the range of likely \av\ is significantly better constrained when all three bands are included.  There are also fewer good models in the fitting with all three bands, indicating a stronger constraint on model parameters.  Thus, we simultaneously fit the 1D radial profiles of three FORCAST bands (19.7, 31.5, and 37.1 \micron) with a 20\arcsec\ strip to constrain the 1D profile.

\subsubsection{Combined IMPRO and SED Fitting}
\label{sec:comb_fitting}
\begin{figure*}[htbp!]
  \centering
  \includegraphics[width=\textwidth]{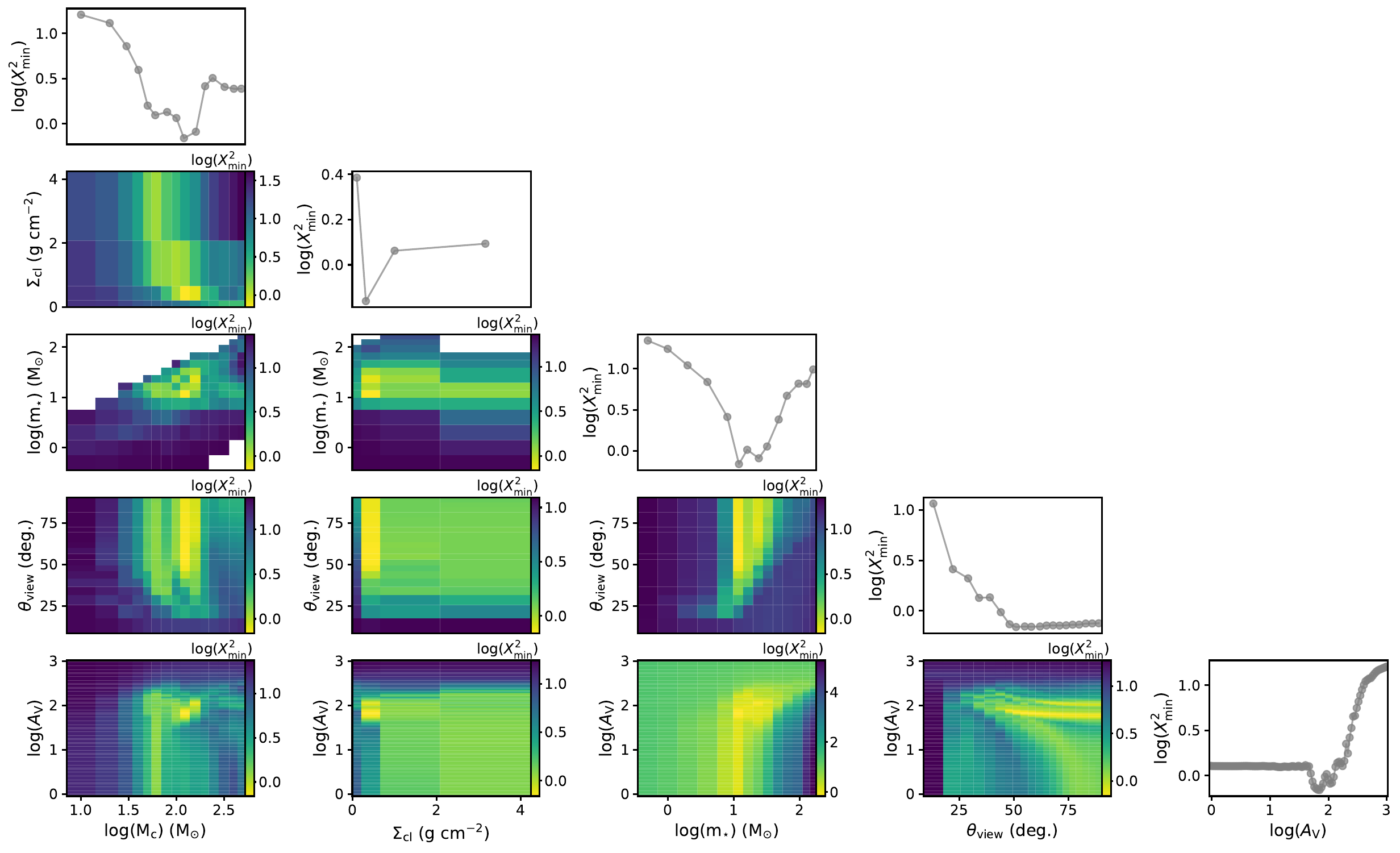}
  \caption{Distributions of the \chised\ marginalized by model parameters for Cep A, G35.2N, and a synthetic ZT18 model.  Each panel shows the distribution of the lowest \chised\ at the given combination of model parameters.  All models, instead of only the good models, are shown in these distributions to highlight the overall \chisq\ distributions.  The gray scatter plots show the minimum \chisq\ as a function of the model parameter in the same column.  The complete figure set (3 figures) is available in the online journal.}
  \label{fig:chi2_sed_summary}
\end{figure*}

As described in Section\,\ref{sec:method}, the \chioned\ is averaged with \chised\ to derive a combined \chicomb\ to assess the goodness of the IMPRO fitting of the 1D intensity profile along the outflow axis and the SED fitting. The distributions of the \chised\ and \chioned\ marginalized by model parameters demonstrate the performance of each method (Figures\,\ref{fig:chi2_sed_summary} and \ref{fig:chi2_1d_summary}).  The minimum \chisq\ with respect to each model parameter is also shown.  The model parameters are not independent, as shown in the distributions.  Thus, the 1D minimum \chisq\ profiles demonstrate the constraining power by the fitting method but cannot indicate the best-fitting parameters.  

When only fitting the SED, \mcore\ and \mstar\ are the most constrained compared to other parameters.  The \mcore\ and \sigmacl\ values appear negatively correlated.  Both \mcore\ and \sigmacl\ contribute to the total dust mass probed by the SED at far-infrared and sub-mm wavelengths, resulting in the degeneracy of these two parameters.  However, the SED fitting yields tighter constraints on \mcore\ and \mstar, effectively breaking the degeneracy between \mcore\ and \sigmacl.  The \view\ and \av, on the other hand, remain poorly constrained via SED fitting.  Their 1D minimum \chisq\ profiles suggest that only upper or lower bounds can be restricted by the SED fitting.  For any given \sigmacl, a wide range of \view\ produces similar quality of fits.  For \mcore\ and \mstar, \view\ appears to be better characterized, although still exhibiting significant degeneracy. The values of \av\ show positive correlation with \mstar, because the more evolved sources (higher \mstar) would have more near-IR emission due to less material in the surrounding envelope, which requires a higher foreground \av\ to produce the same near-IR fluxes.  The \chised\ distribution of \av\ and \view\ shows multiple strips of negative correlation.  A source with a higher \view\ would have a redder SED, while \av\ has a similar effect on the SED.  

\renewcommand{\thefigure}{\arabic{figure} (Cont.)}
\addtocounter{figure}{-1}

\begin{figure*}[htbp!]
  \centering
  \includegraphics[width=\textwidth]{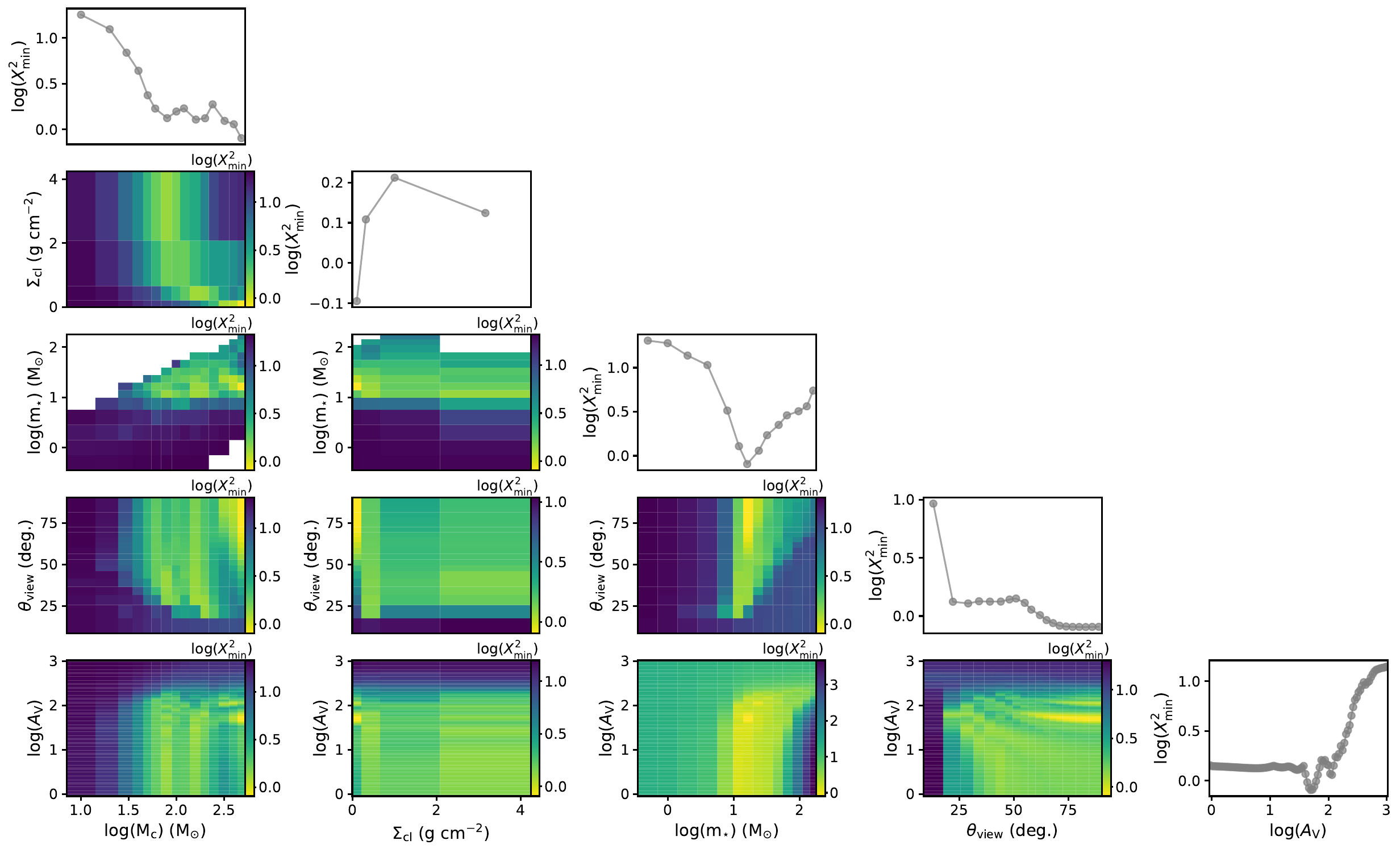}
  \caption{Distributions of the \chised\ marginalized by model parameters for G35.2N.}
\end{figure*}

\addtocounter{figure}{-1}

\begin{figure*}[htbp!]
  \centering
  \includegraphics[width=\textwidth]{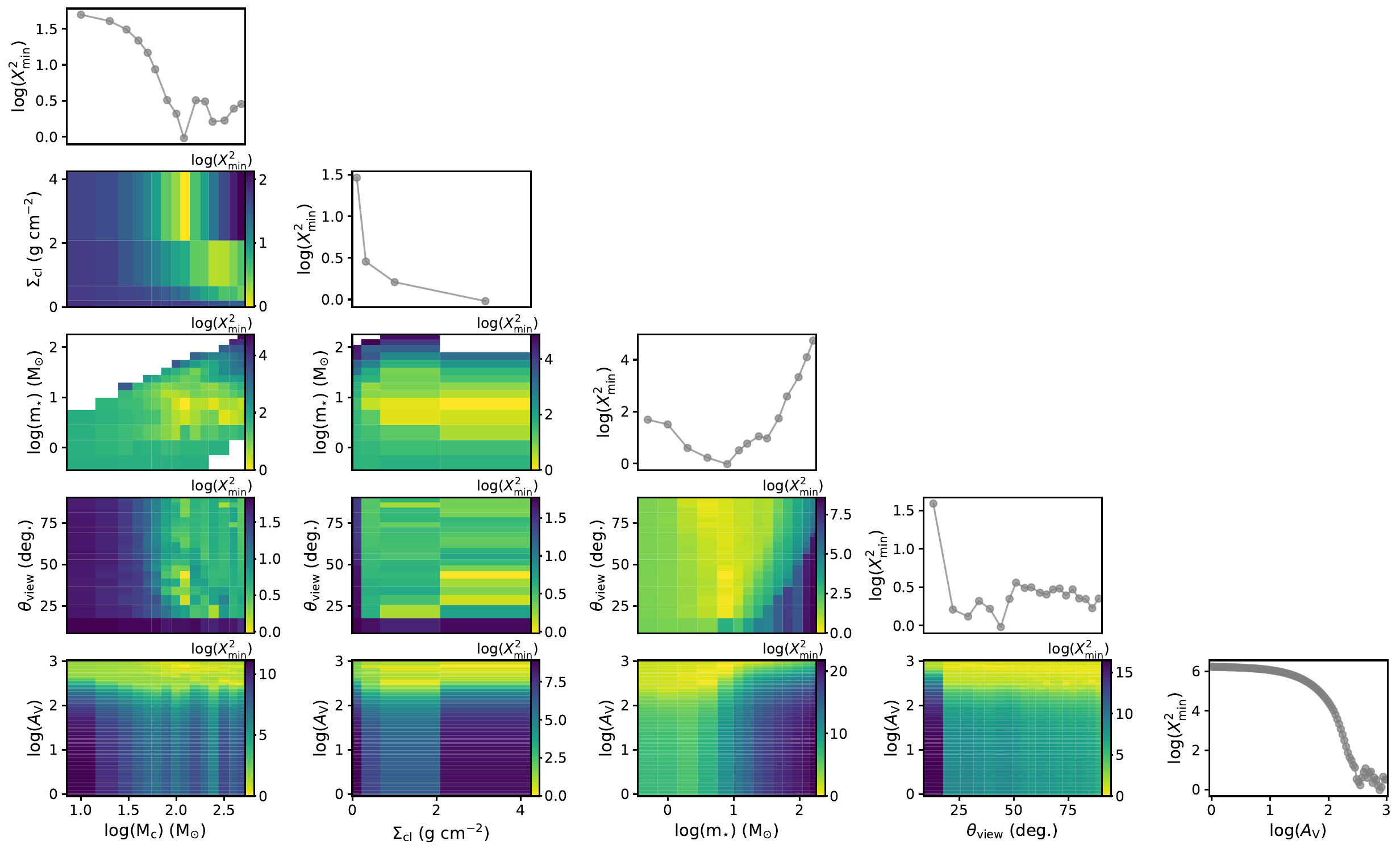}
  \caption{Distribution of the \chised\ marginalized by model parameters for the synthetic ZT18 model.}
\end{figure*}

\renewcommand{\thefigure}{\arabic{figure}}

The distribution of \chioned\ exemplifies the constraining power of IMPRO fitting that is distinct from that resulting from SED fitting (Figure\,\ref{fig:chi2_1d_summary}). The good models have a narrow range of \view\ in all distributions marginalized by \view.  The 1D minimum \chisq\ also indicates a clear preferred value.  The best-fitting parameters qualitatively agree with those derived from SED fitting, except for \sigmacl.  IMPRO fitting prefers a high \sigmacl, whereas SED fitting prefers a lower \sigmacl, although the \chised\ varies relatively less with \sigmacl\ compared to other parameters.  Overall, the IMPRO fitting provides an improved estimate on \view, while produces similar constraints on \mcore\ and \mstar.  For \av, the difference between the SED and the IMPRO fittings is marginal.

\begin{figure*}[htbp!]
  \centering
  \includegraphics[width=\textwidth]{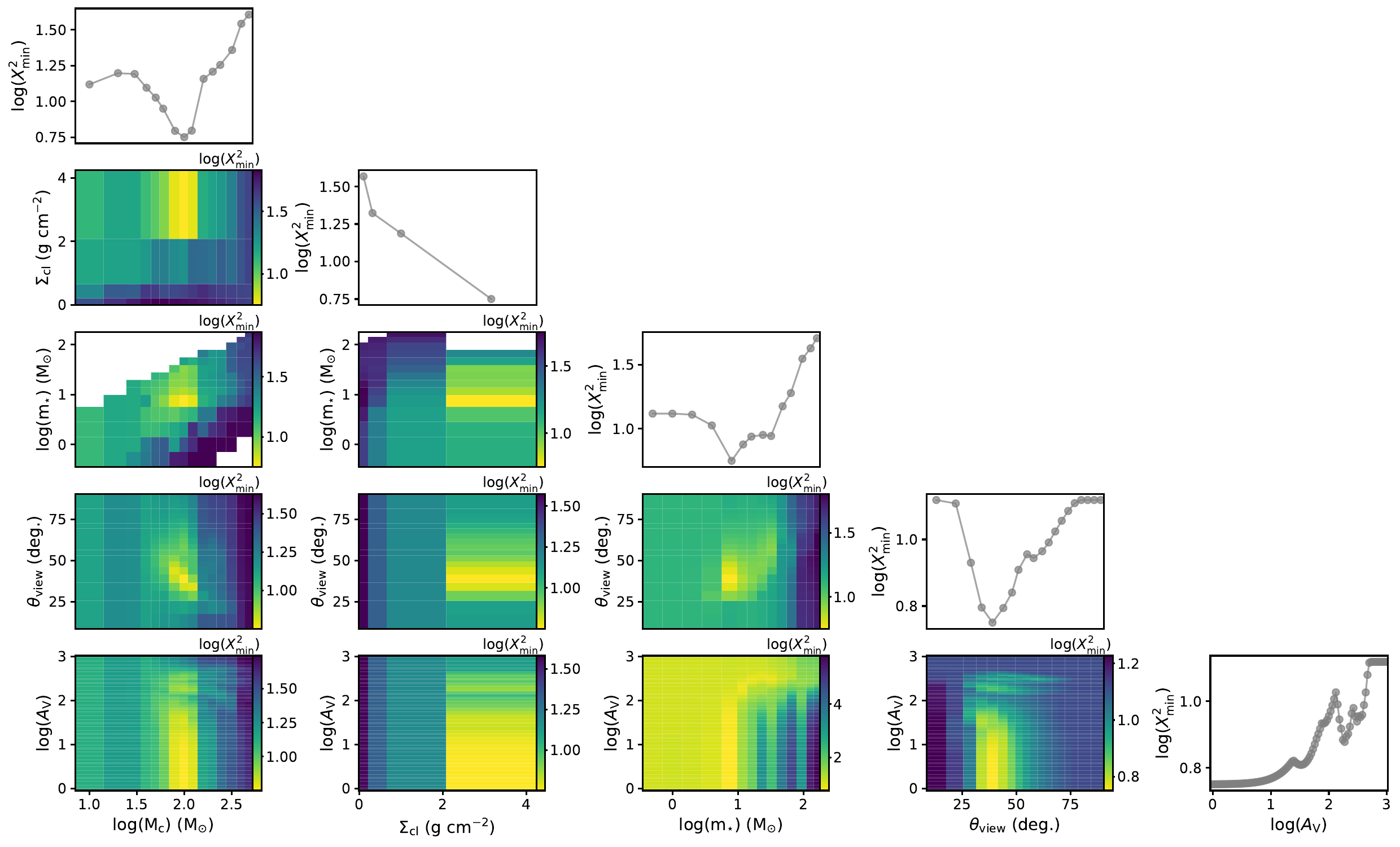}
  \caption{Distribution of the \chioned\ marginalized by model parameters for Cep A, G35.2N, and a synthetic ZT18 model.  The IMPRO fitting uses all three bands, 37.1, 31.5, and 19.7 \micron. The method and the legends are similar to those in Figure\,\ref{fig:chi2_sed_summary}.  The complete figure set (3 figures) is available in the online journal.}
  \label{fig:chi2_1d_summary}
\end{figure*}

\renewcommand{\thefigure}{\arabic{figure} (Cont.)}
\addtocounter{figure}{-1}

\begin{figure*}[htbp!]
  \centering
  \includegraphics[width=\textwidth]{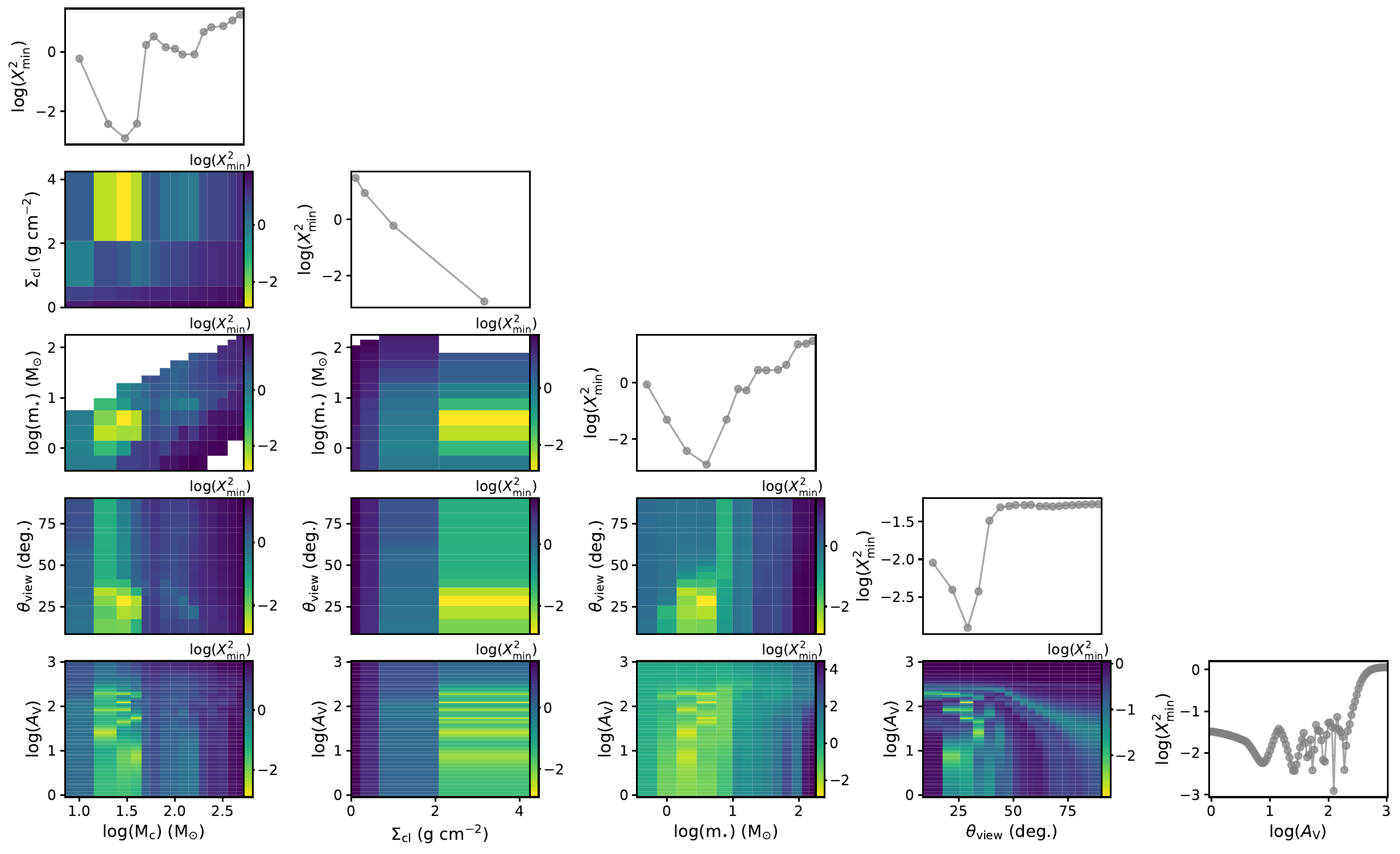}
  \caption{Distributions of the \chioned\ marginalized by model parameters for G35.2N.}
\end{figure*}

\addtocounter{figure}{-1}

\begin{figure*}[htbp!]
  \centering
  \includegraphics[width=\textwidth]{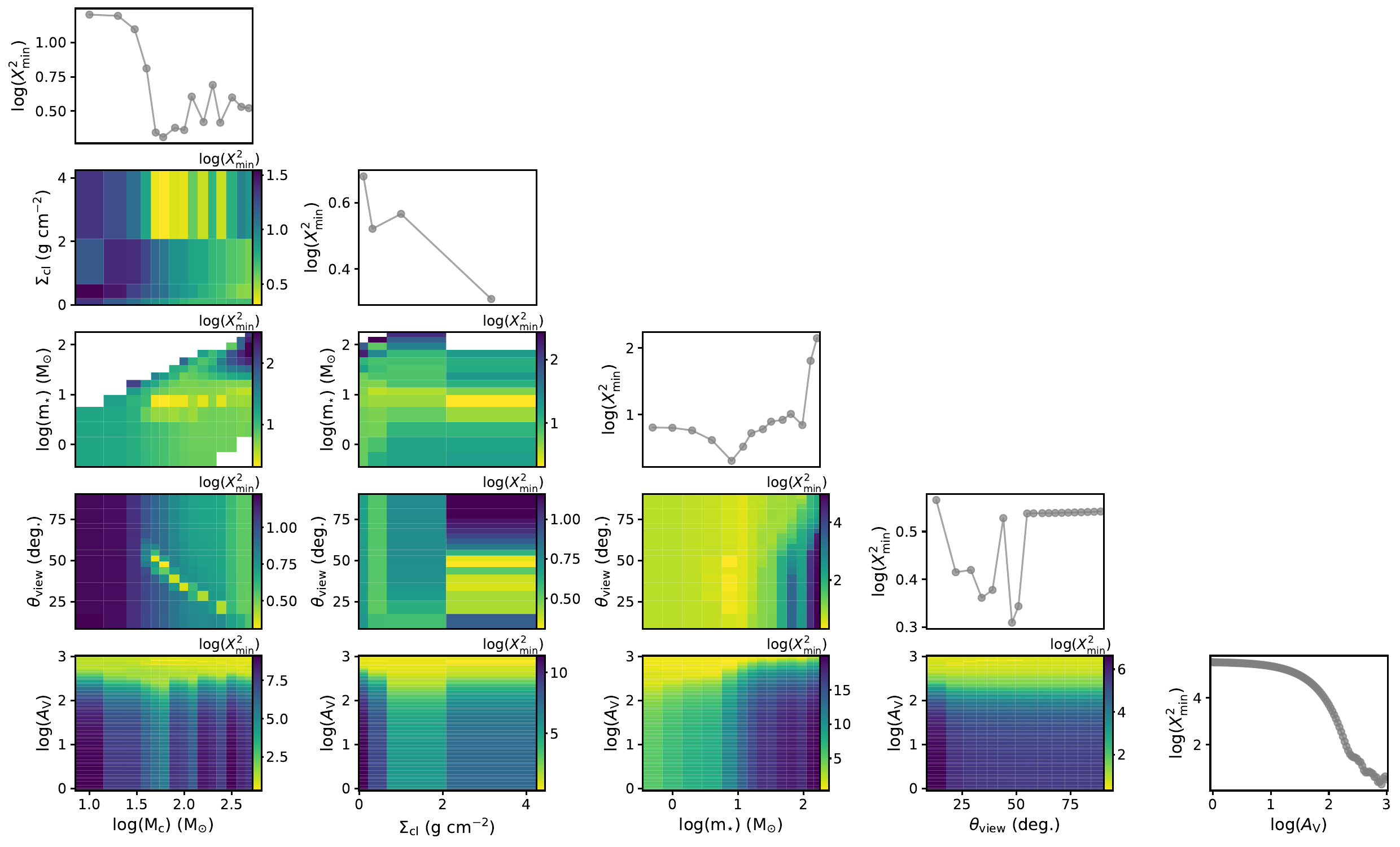}
  \caption{Distribution of the \chioned\ marginalized by model parameters for the synthetic ZT18 model.}
\end{figure*}

\renewcommand{\thefigure}{\arabic{figure}}

\begin{figure*}[htbp!]
  \centering
  \includegraphics[width=\textwidth]{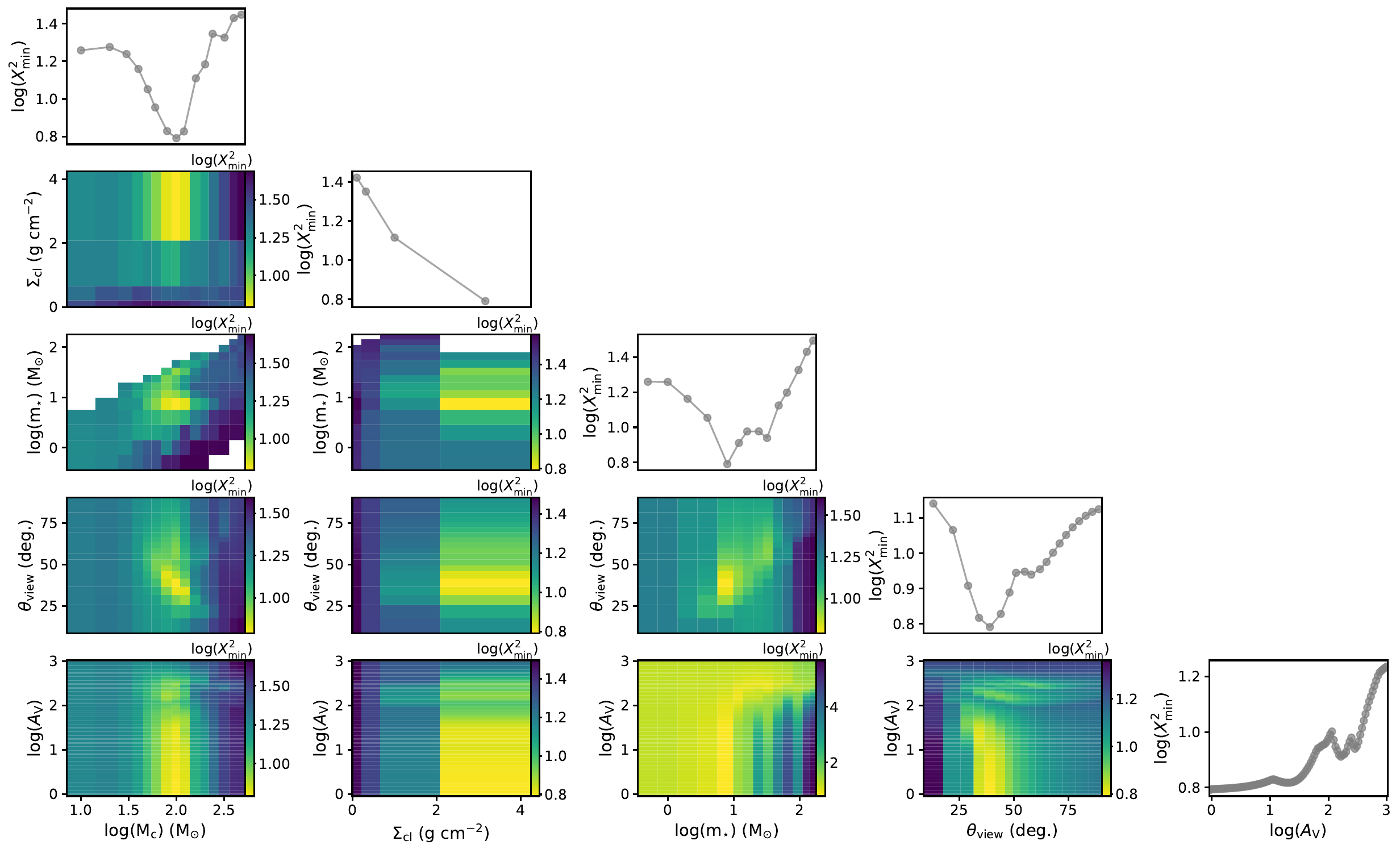}
  \caption{Distribution of the \chicomb\ marginalized by model parameters for Cep A, G35.2N, and a synthetic ZT18 model.  The IMPRO fitting uses all three bands, 37.1, 31.5, and 19.7 \micron.  The method and the legends are similar in Figure\,\ref{fig:chi2_sed_summary}.  The complete figure set (3 figures) is available in the online journal.}
  \label{fig:chi2_comb_summary}
\end{figure*}

Figure\,\ref{fig:chi2_comb_summary} shows the marginalized \chicomb\ distribution of the combined IMPRO and SED fitting for Cep A.  As described in Section\,\ref{sec:method}, the \chicomb\ is the average of \chised\ and \chioned, giving them an equal weight.  The distributions, therefore, present the characteristics of each fitting method discussed in the previous paragraphs.  Most model parameters have local minima in these marginalized \chicomb\ distributions, while \av\ has no clear $\chi^2$ minimum, as a result of weak constraints from both the SED and IMPRO fitting methods.  The minimum \chicomb\ is higher than either the minimum \chised\ or the minimum \chioned, suggesting that this fitting method, despite the improved constraints in \av\ and \view, cannot fully capture the comprehensive properties of the massive protostellar cores.  The minimum \chicomb\ is 6.18, also hinting at some limitations of the ZT18 model grid.

\renewcommand{\thefigure}{\arabic{figure} (Cont.)}
\addtocounter{figure}{-1}

\begin{figure*}[htbp!]
  \centering
  \includegraphics[width=\textwidth]{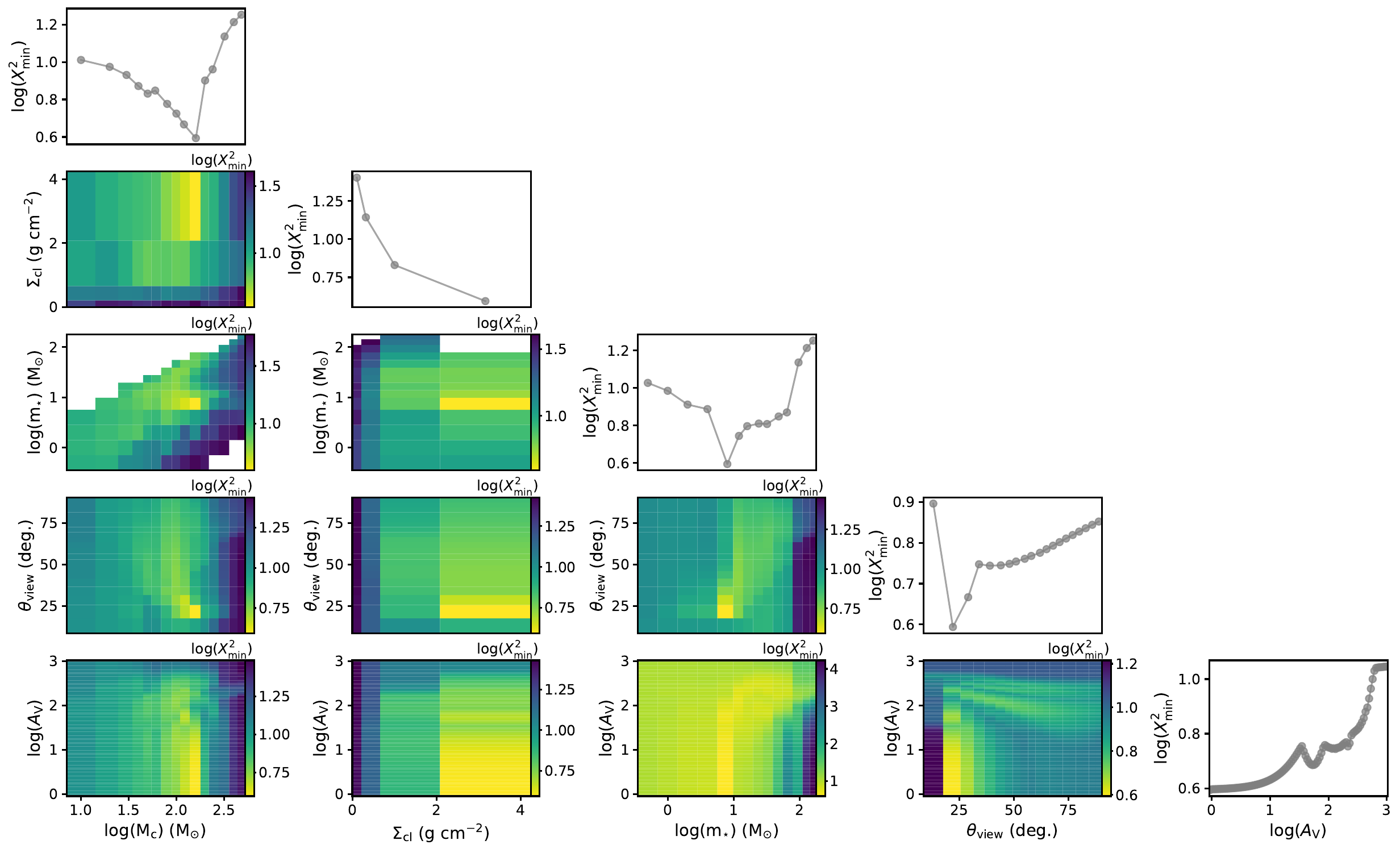}
  \caption{Distributions of the \chicomb\ marginalized by model parameters for G35.2N.}
\end{figure*}

\addtocounter{figure}{-1}

\begin{figure*}[htbp!]
  \centering
  \includegraphics[width=\textwidth]{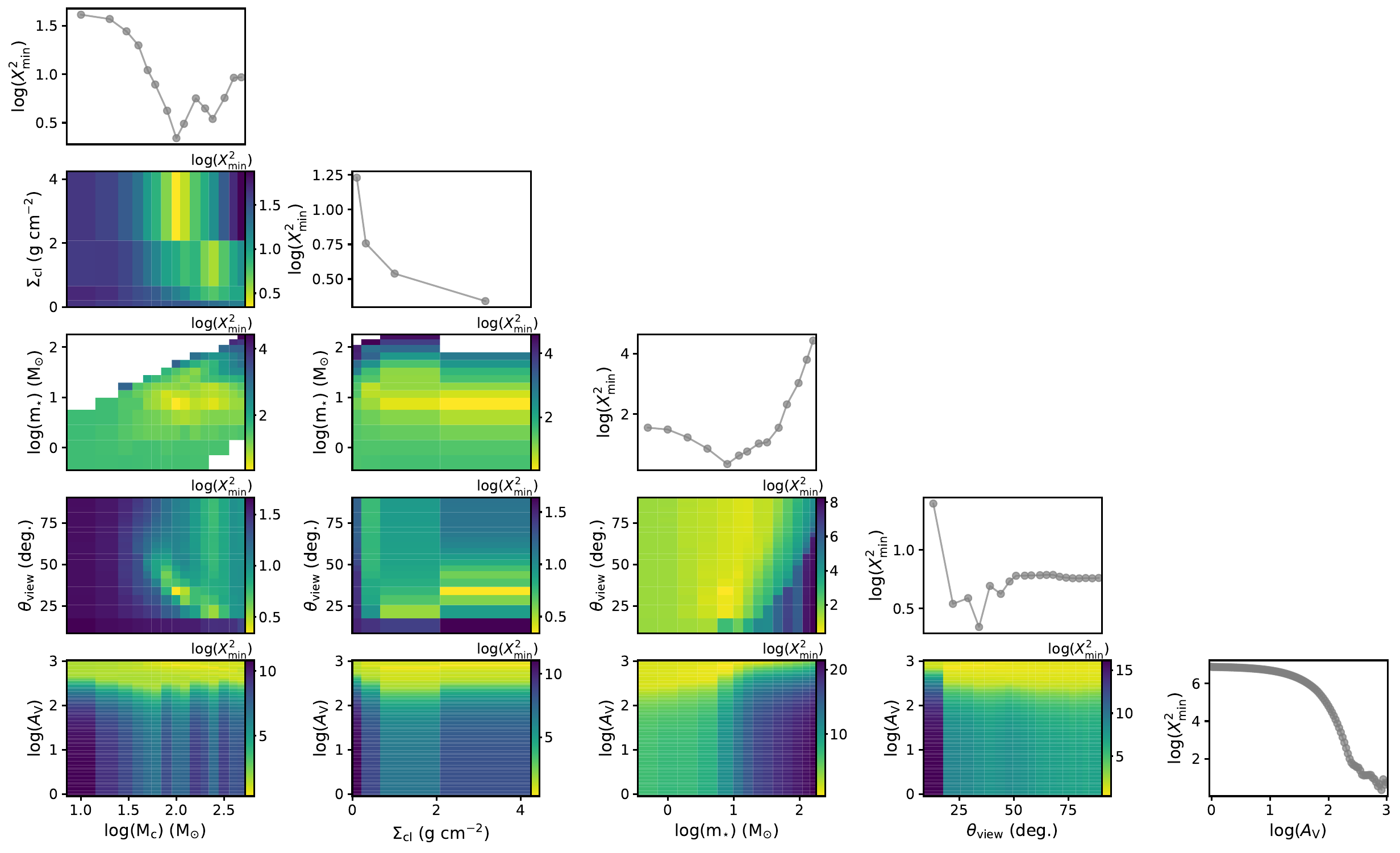}
  \caption{Distribution of the \chicomb\ marginalized by model parameters for the synthetic ZT18 model.}
\end{figure*}

\renewcommand{\thefigure}{\arabic{figure}}

\subsubsection{Best-fitting Parameters of Cep A}
\label{sec:bestfit_cepA}

\begin{figure*}[htbp!]
  \centering
  \includegraphics[width=\textwidth]{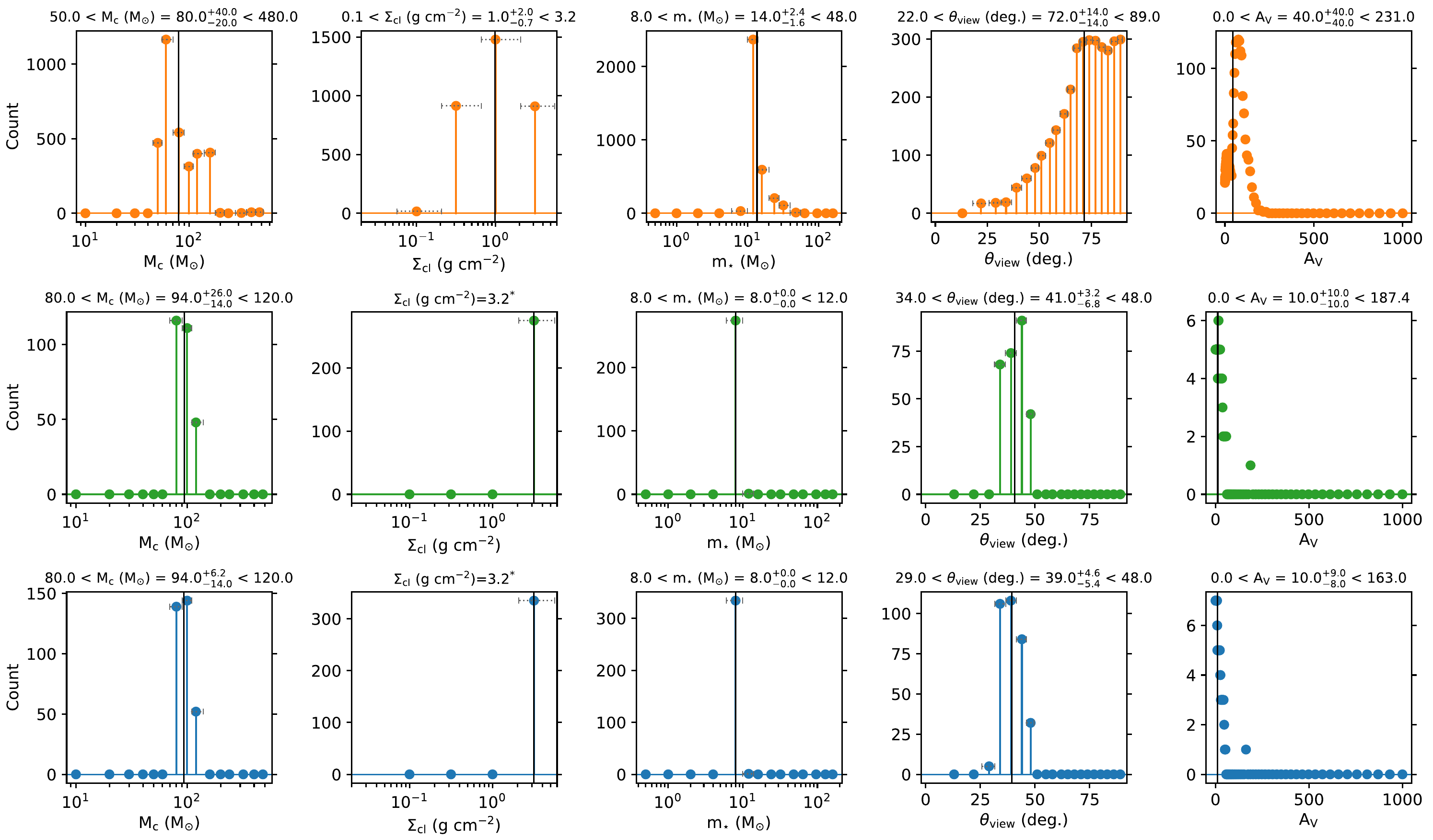}
  \caption{Distribution of the fitted parameters for the good models for Cep A, G35.2N (with a minimum of three and five data points), and a synthetic ZT18 model.  The good models are selected by $\chi^2_\text{min} < \chi^2 < \chi^2_\text{min}+2$. The row shows the distribution of $\chi^2_{\rm SED}$, $\chi^2_{\rm IMPRO}$, and $\chi^2_{\rm SED+IMPRO}$ from top to bottom.  The mean value and the range of the model parameters are shown at the top of each panel.  An asterisk indicates that all models have the same value.  The minimum \chised, \chioned, and \chicomb\ are 0.69, 5.63, and 6.18, respectively.  Note that the models with the minimum \chised\ and \chioned\ may not be the same model.  Similar to Figure\,\ref{fig:width_comp}, the gray horizontal bars show the range of the parameter value that is represented by the particular radiative transfer model in the ZT18 model grid.  The complete figure set (4 figures) is available in the online journal.
  }
  \label{fig:model_parameters}
\end{figure*}

Figure\,\ref{fig:model_parameters} shows the distribution of the model parameters among the good models for the fitting using the SED-only, IMPRO-only, and SED+IMPRO.  Similar to the discussion in Section\,\ref{sec:comb_fitting}, the SED fitting suggests a well-fitted \mstar\ with moderately constrained \mcore.  The \sigmacl, \view, and \av\ are not well-constrained.  The \mcore\ of the good models ranges over $\sim50-400$\%\ from the geometric mean value of 79.7 \msun.  The good models show a broad distribution of \sigmacl, peaking at the 1.0~g~cm$^{-2}$ bin.  Moreover, \view\ is hardly constrained, with a broad distribution ranging from 22$^{\circ}$ to $\sim$90$^{\circ}$.  The \av\ distribution has bimodal peaks, suggesting likely degeneracies among the model parameters.  On the other hand, IMPRO fitting constrains \view\ within $\sim 10^{\circ}$ of the mean value, a drastic improvement from the results of the SED fitting.  The fitted \mcore\ and \mstar\ have a similar mean value as that of the SED fitting with a narrower range.  All the good models have the same \sigmacl\ and almost the same \mstar.  IMPRO fitting also constrains \av\ with a more coherent distribution.  The bimodal distribution seen in the \av\ fitted by SED becomes a relatively uniform distribution with an upper limit, suggesting that the IMPRO fitting prefers a lower \av, while a specific \av\ cannot be determined, possibly due to the limited number of bands included in the fitting.  IMPRO fitting alone can reproduce the mean values of the model parameters found from SED fitting and yield Gaussian-like distributions, suggesting a better fitting performance.  Figure\,\ref{fig:model_parameters} (bottom) shows the model parameters estimated with the combined SED + IMPRO fitting.  The mean values of model parameters are similar to those derived with IMPRO fitting, suggesting that the 1D intensity profiles have a stronger influence on the resulting \chicomb.

\begin{table*}[htbp!]
  \caption{The best-fitting model parameters of Cep A and G35.2N \label{tbl:best_params}}
  \centering
  \begin{tabular}{l|rclrclrcl}
    \toprule
    \multicolumn{1}{r}{\quad} & \multicolumn{9}{c}{Fitting method} \\
    \cline{2-10}
    Cep A & \multicolumn{3}{c}{\chised} & \multicolumn{3}{c}{\chioned} & \multicolumn{3}{c}{\chicomb} \\
    \midrule
    \mcore\ (\msun)         & 50$<$   & 80$^{+40}_{-20}$    &   $<$480  & 80$<$       & 94$^{+26}_{-14}$        & $<$120      & 80$<$       & 94$^{+6.2}_{-14}$       & $<$120      \\
    \sigmacl\ (g cm$^{-2}$) & 0.1$<$  & 1.0$^{+2}_{-0.7}$   &   $<$3.2  & 3.2$^\star$ & 3.2$^\star$             & 3.2$^\star$ & 3.2$^\star$ & 3.2$^\star$             & 3.2$^\star$ \\
    \mstar\ (\msun)         & 8$<$    & 14$^{+2.4}_{-1.6}$  &   $<$48   & 8$<$        & 8$^{+0}_{-0}$$^\dagger$ & $<$12       & 8$<$        & 8$^{+0}_{-0}$$^\dagger$ & $<$12       \\
    \view\ ($^{\circ}$)     & 22$<$   & 72$^{+14}_{-14}$    &   $<$89   & 34$<$       & 41$^{+3.2}_{-6.8}$      & $<$48       & 29$<$       & 39$^{+4.6}_{-5.4}$      & $<$48       \\
    \av\ (mag)              & 0$<$    & 40$^{+40}_{-40}$    &   $<$231  & 0$<$        & 10$^{+10}_{-10}$        & $<$187      & 0$<$        & 10$^{+9}_{-8}$          & $<$163      \\
    \midrule
    G35.2N & \multicolumn{3}{c}{\chised} & \multicolumn{3}{c}{\chioned} & \multicolumn{3}{c}{\chicomb} \\
    \midrule
    \mcore\ (\msun)         & 60$<$   & 170$^{+68}_{-52}$   &   $<$480  & 40$<$       & 74$^{+46}_{-34}$        & $<$160      & 100$<$      & 120$^{+44}_{-16}$       & $<$160      \\
    \sigmacl\ (g cm$^{-2}$) & 0.1$<$  & 0.4$^{+0.6}_{-0.0}$$\dagger$ & $<$3.2 & 1.0$<$& 2.0$^{+1}_{-0.9}$       & $<$3.2      & 3.2$^\star$ & 3.2$^\star$             & 3.2$^\star$ \\
    \mstar\ (\msun)         & 12$<$   & 16$^{+8}_{-4}$      &   $<$48   & 8$<$        & 9.3$^{+2.7}_{-1.3}$     & $<$32       & 8$<$        & 8.6$^{+3.4}_{-0.64}$    & $<12$       \\
    \view\ ($^{\circ}$)     & 22$<$   & 65$^{+18}_{-17}$    &   $<$89   & 22$<$       & 41$^{+14}_{-12}$        & $<$74       & 22$<$       & 30$^{+4}_{-8}$          & $<$58       \\
    \av\ (mag)              & 0$<$    & 20$^{+20}_{-20}$    &   $<$187  & 0$<$        & 40$^{+60}_{-40}$        & $<$266      & 0$<$        & 30$^{+50}_{-30}$        & $<$231      \\
    \bottomrule
    \multicolumn{10}{p{5in}}{\textsc{Note}: The 16th and 84th percentile values as well as the full range of the fitted parameters are listed.  The percentiles do not statistically represent the 1-$\sigma$ confidence interval due to the ill-defined $\chi^2$ statistics, but they provide another perspective on how the model parameters distributed within the good models. The significant figures are chosen to be conservative given the statistical ambiguities (Section\,\ref{sec:method}).  The dagger indicates the cases where most models have the same values, resulting in zero in the chosen percentiles.  The asterisk indicates that all the good models have the same value.}
  \end{tabular}
\end{table*}

Figure\,\ref{fig:best_fitting_models} shows the models with the lowest $\chi^2$ fitted by the SED, IMPRO, and SED+IMPRO methods.  As expected, the best-fitting model for the SED fitting reproduces the observed SED.  The best-fitting SED model produces 1D profiles largely inconsistent with the observed profiles.  The brightness in the SED model peaks at the source position, while the observed profiles have their brightness peak at an offset position.  The 1D profiles from the SED model also underestimate the brightness at positive offsets corresponding to the blue-shifted outflow cavity.  On the other hand, the best IMPRO fitted model reproduces the observed 1D profiles except for the exact shape of the 1D profile at 19.7 \micron.  The synthetic 1D profile at 19.7 \micron\ has a smaller contrast between the two outflow cavities compared to that shown in the observed profile.  However, both profiles peak at a similar position.  The best-fitting IMPRO model yields a slightly worse SED model, which reproduces the far-infrared fluxes but underestimates the mid-infrared fluxes.  While the IMPRO fitting considers the mid-IR fluxes (19--37 \micron), the fluxes used in the IMPRO and SED are extracted from different apertures.  The IMPRO uses a 20\arcsec\ strip with a length corresponding to the model core radius, resulting in a $\sim20\arcsec\times30\arcsec$ region.  On the other hand, the SED uses a circular aperture with a 36\farcs{25} radius.  The underestimate in the mid-IR fluxes by the IMPRO fitting suggests that some features in the core-to-clump transition regime are not fully captured by the ZT18 model grid and our modeling treatment.  The best model in the combined fitting is the same model as the best-fitting IMPRO model. Interestingly, both the best IMPRO and the combined model have \av=0.  As one of the nearest massive star-forming cores, Cep A could have a lower \av, but \av=0 is likely unreasonable.  The mismatch in the extracting areas between the 1D profile and the SED may also contribute to the lower \av\ in the best-fitting IMPRO and SED+IMPRO models.  Thus, we recommend using the mean values and the associated uncertainties and ranges shown in Table\,\ref{tbl:best_params} instead of the best-fitting models to estimate the physical properties of Cep A as well as the sources fitted using the IMPRO+SED approach.

The drastic difference in the 1D profiles fitted by the SED and IMPRO methods comes from the fitted core properties.  The SED model (\zt{120.0}{0.316}{12.0}{51.0} and \av=70.5) prefers a larger core ($R_c = 0.3$ pc), while the 1D profile model (\zt{100.0}{3.16}{8.0}{29.0} and \av=0.0) converges on a smaller core ($R_c = 0.05$ pc). Interestingly, the best-fitting IMPRO model fits the 1D profiles at 31.5 and 37.1 \micron\ but underestimates the photometric fluxes at those bands on the SED.  The 1D profile captures the brightness variation along the outflow axis, but ignores the shape of the outflow cavities, such as their opening angle.  A larger opening angle would lead to a higher flux at mid-infrared wavelengths \citep[e.g.,][]{2017ApJ...835..259Y}.  Thus, the shape of outflow cavities may provide additional constraints for further improvements of the modeling. 

We further compare the fitted model parameters with similar parameters derived from other observational studies.  Introducing the 1D profile significantly improves the characterization of the \view\ to be 39.4$^\circ$$^{+4.6^\circ}_{-5.4^\circ}$ (Table\,\ref{tbl:best_params}) changed from 71.6$^\circ$$^{+14.4^\circ}_{-13.6^\circ}$ fitted with the SED.  The inclination of the disk in Cep A is measured as 62$^\circ$ and 64$^\circ$ using dust and \methylcyanide\ emission as well as the kinematics of \methanol\ maser spots, respectively \citep{2005Natur.437..109P,2017AA...603A..94S}.  If the outflow is perpendicular to the disk midplane, the \view\ would be 28$^\circ$ and 26$^\circ$.  Our combined fitting method estimates the \view\ slightly greater than the value inferred from disk inclination.  Considering the uncertainty in the measurement of disk inclination angle ($12^\circ\pm9^\circ$; \citealt{2017AA...603A..94S}), our best-fitting value is similar to that inferred from disk inclination.

The studies on the disk of Cep A also estimate the protostellar mass with dynamical analysis.  We focus on the measurements specifically attributed to HW2, the primary outflow-driven source.  \citet{2017AA...603A..94S} derived two masses of 5.5 and 10 \msun\ for HW2.  The former reproduces the observed kinematics but only accounts for 10\%\ of the observed \lbol, while the latter matches the \lbol\ but would indicate that magnetic fields reduce the observed maser velocities from the true Keplerian motion.  The magnetically reduced velocities are supported by the detection of polarized dust and line emission in Cep A.  Recently, a modeling study of the NH$_3$ kinematics suggests a central mass of 16 \msun\ and a Keplerian disk of $\sim200$ au \citep{2025AA...697A.206S}.  They tested the models with 12 and 16 \msun\ and found the model with a higher mass better reproduces the observation.  In comparison, our combined fitting suggests a best-fitting \mstar\ of 8.0 \msun\ with the next mass sampling at 12 \msun.  Our estimate is on the lower end of the available mass measurements in literature and can be about a factor of 2 lower than that derived from high-resolution gas kinematic.  The mass estimated from the SED-only fitting is higher, 13.6$^{+2.4}_{-1.6}$ \msun, and closer to the other mass measurements.  This difference highlights the need for follow-up studies to test the balance between the SED and IMPRO fittings using observational data.  This modest inconsistency may be rooted in the stellar evolutionary track and the outflow opening mechanism in the ZT18 model grid.

Other model parameters, \mcore, \sigmacl, and \av, cannot be compared with observations directly.  The \mcore\ and \sigmacl\ are the properties of the \textit{initial} core, while observations only probe the core mass and surface density at the current time.  However, we can use the current core mass to set a lower limit on the mass of the initial core.  \citet{1991ApJ...374..169M} found a hierarchical structure consisting of an extended (0.5 pc$\times$0.85 pc) core with a mass of 250--400 \msun, a concentration (0.5 pc$\times$0.25 pc) with a mass of 200--300 \msun, and a hot core ($\sim$0.14 pc) with a mass greater than 70 \msun.  The hot core size, $\sim40\arcsec$, is similar to the size of the best-fitting core model with $\sim0.05$ pc.  Compared to the SED+IMPRO results, the fitted \mcore, 93.8$^{+6.2}_{-13.8} $\msun, is consistent with the lower limit set by the mass of the hot core.

We further compare the envelope mass (\menv) in the ZT18 models, tracing the current mass.  The distribution of the \menv\ is 16\%\ lower than the initial core mass with a geometric mean value of 78.7 \msun\ (Figure\,\ref{fig:current_mass}, left).  The estimated \menv\ remains consistent with the other independent observational constraints. Overall, our combined fitting of Cep A constrains the model parameters that are consistent with other independent measurements, especially the \view, suggesting that this fitting pipeline would yield model parameters that can characterize massive protostellar cores.

\renewcommand{\thefigure}{\arabic{figure} (Cont.)}
\addtocounter{figure}{-1}

\begin{figure*}[htbp!]
  \centering
  \includegraphics[width=\textwidth]{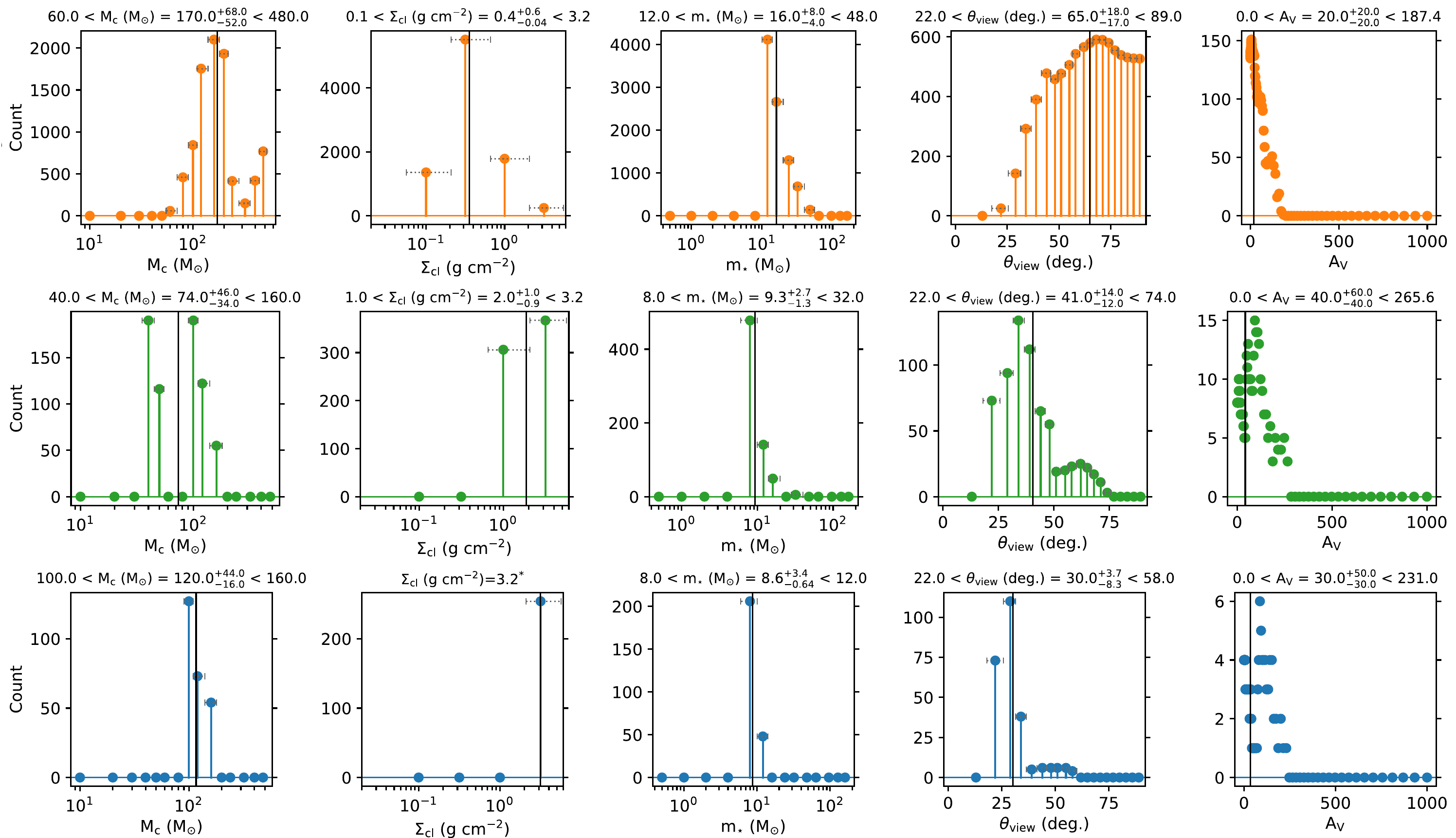}
  \caption{Distribution of the fitted parameters for the good models for G35.2N with at least 3 data points in any of its 1D profiles.  The minimum \chised, \chioned, and \chicomb\ are 0.80, 0.83, and 3.92, respectively.}
\end{figure*}

\addtocounter{figure}{-1}

\begin{figure*}[htbp!]
  \centering
  \includegraphics[width=\textwidth]{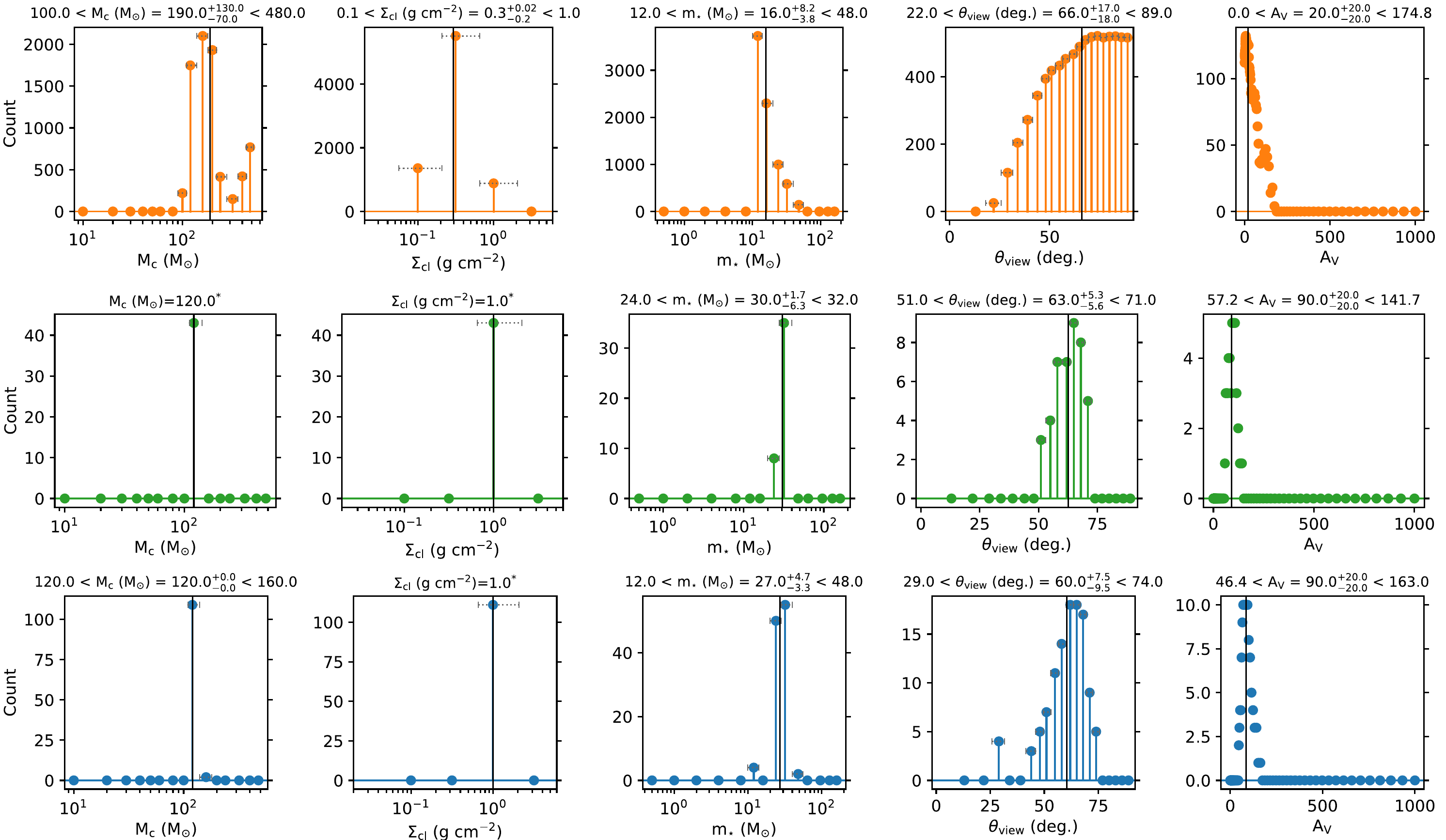}
  \caption{Distribution of the fitted parameters for the good models for G35.2N with at least 5 data points in any of its 1D profiles.  The minimum \chised, \chioned, and \chicomb\ are 0.80, 8.20, and 6.90, respectively.}
\end{figure*}

\addtocounter{figure}{-1}

\begin{figure*}[htbp!]
  \centering
  \includegraphics[width=\textwidth]{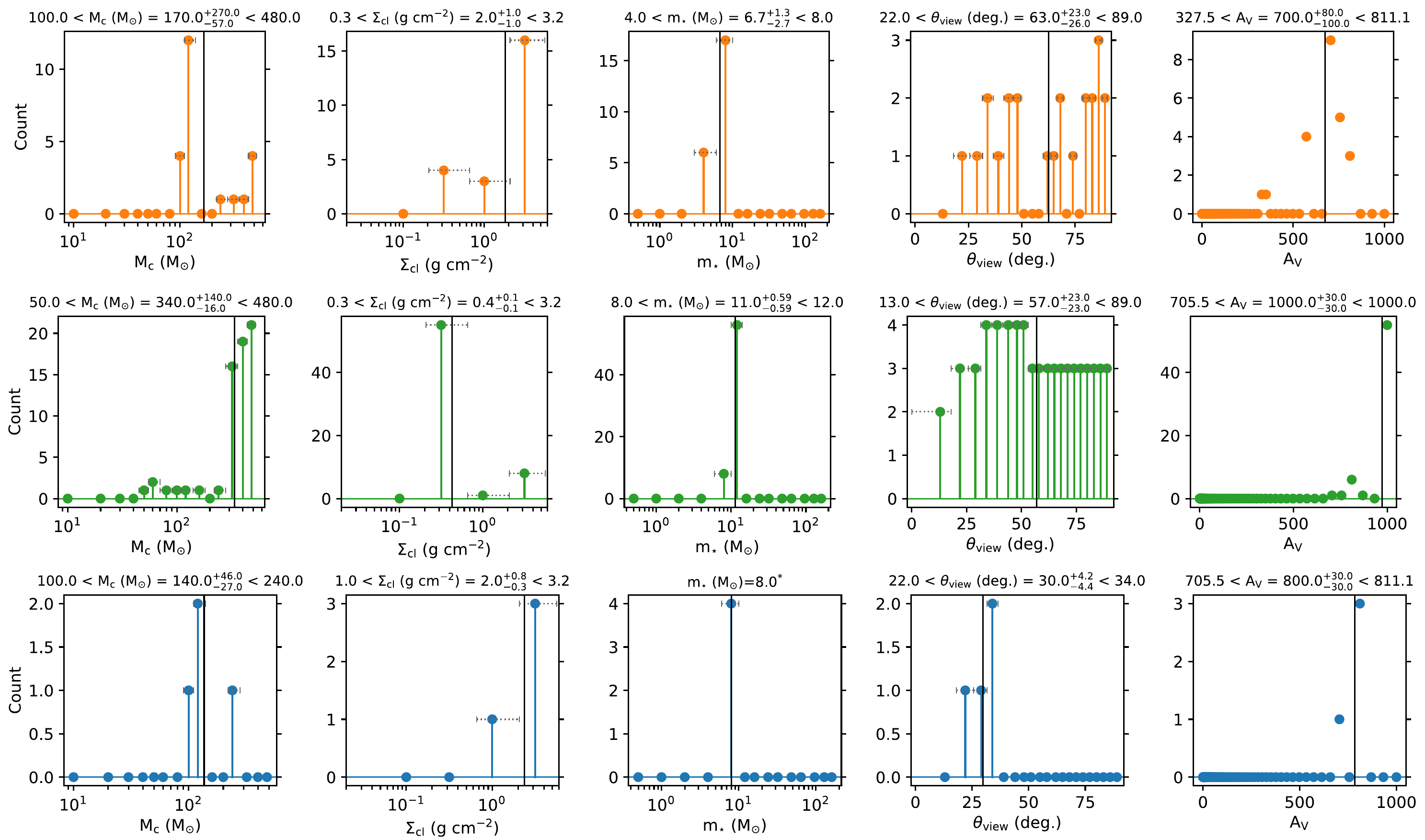}
  \caption{Distribution of the fitted parameters for the good models for the synthetic ZT18 model.  The minimum \chised, \chioned, and \chicomb\ are 0.95, 2.03, and 2.19, respectively.}
\end{figure*}

\renewcommand{\thefigure}{\arabic{figure}}

\begin{figure*}[htbp!]
  \centering
  \includegraphics[width=\textwidth]{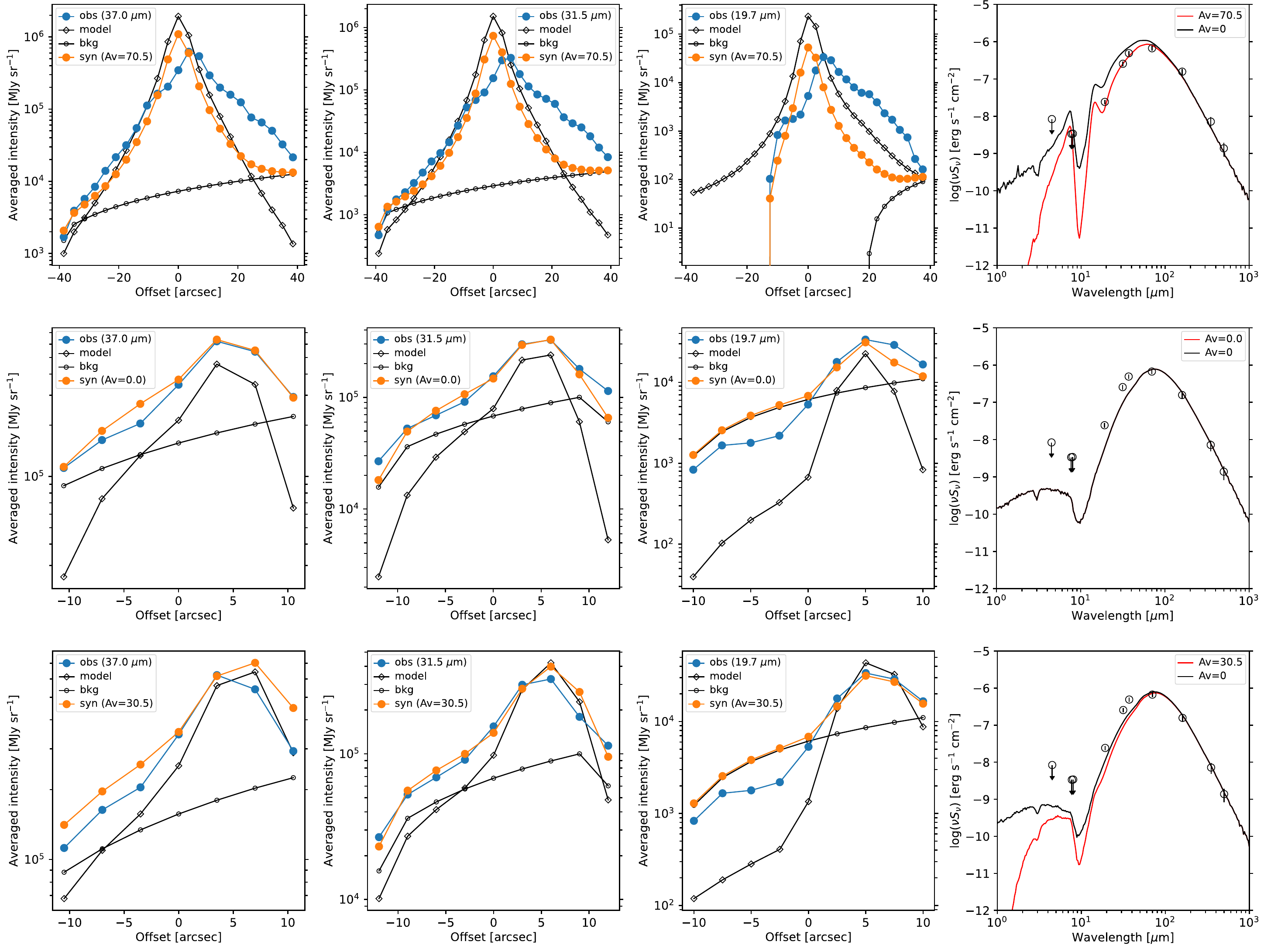}
  \caption{The best-fitting models fitted with the SED, IMPRO, and the combined SED+IMPRO (from top to bottom) for Cep A, G35.2N (with a minimum of three and five data points), and a synthetic ZT18 model.  The 1D profiles at 37.1, 31.5, and 19.7 \micron\ are shown from left to right, together with the SED at the rightmost.  In the panels of the 1D profiles, the blue filled circles and black open circles show the observed 1D profile and the measured background profile from the observed profile.  The orange filled circles and the open diamond show the synthetic 1D profile, including the effect of \av, and the 1D profile in the model.  The extent of the radial profile is determined by the best-fitting core radius, leading to these radial profiles having different extension in offsets.  In the SED panels, the observed photometry is shown in open circles, while the red and black curves show the synthetic SED with and without \av.  The SED values are extracted from a 72\farcs{5} aperture by \citet{2023ApJ...942....7F}.  The complete figure set (4 figures) is available in the online journal.}
  \label{fig:best_fitting_models}
\end{figure*}

\renewcommand{\thefigure}{\arabic{figure} (Cont.)}
\addtocounter{figure}{-1}

\begin{figure*}[htbp!]
  \centering
  \includegraphics[width=\textwidth]{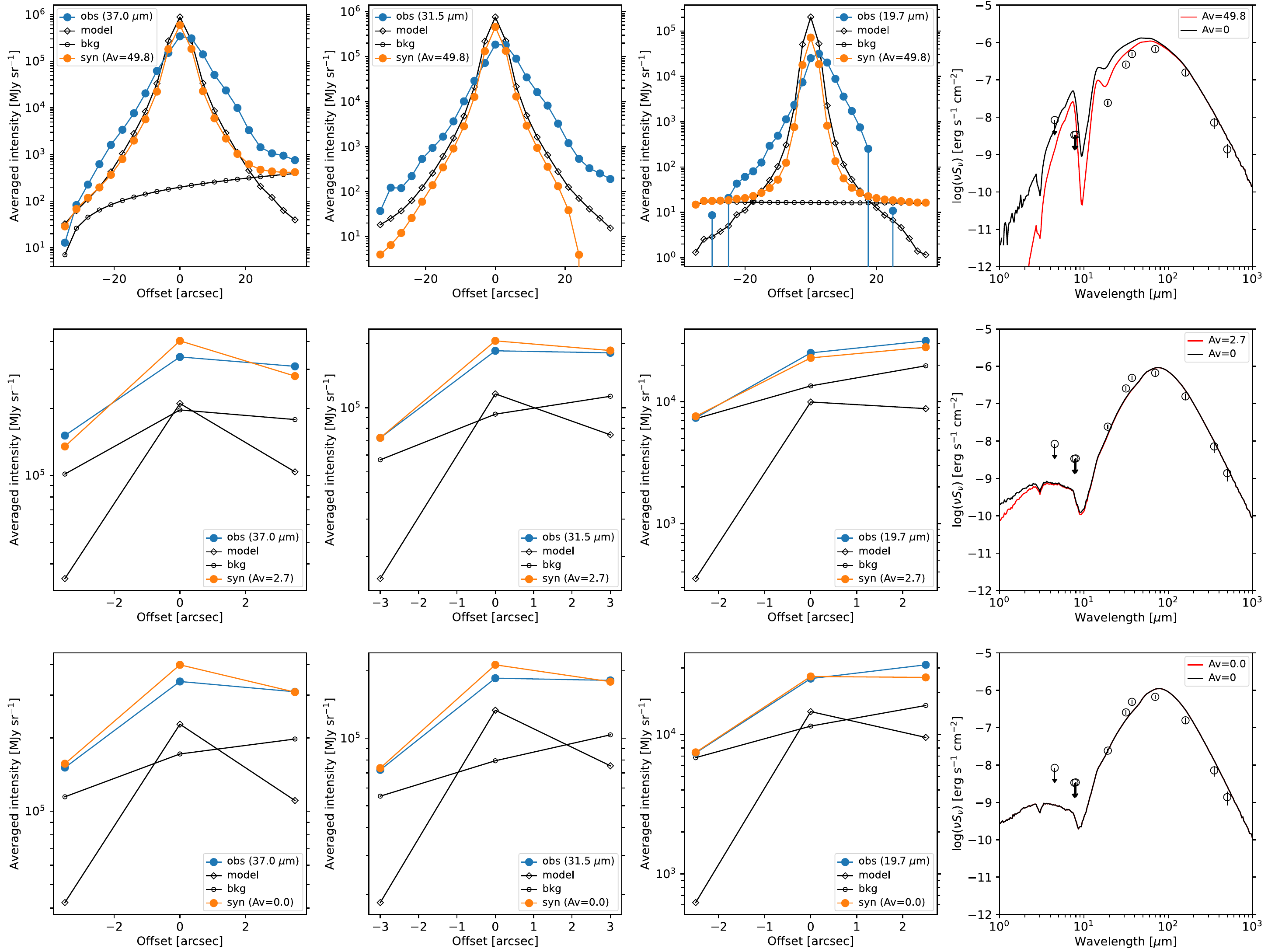}
  \caption{The best-fitting models of G35.2 fitted with the SED, IMPRO, and the combined SED+IMPRO (from top to bottom) among the models with at least 3 data points.  The SED values are extracted from a 37\farcs{5} aperture by \citet{2023ApJ...942....7F}.}
\end{figure*}

\addtocounter{figure}{-1}

\begin{figure*}[htbp!]
  \centering
  \includegraphics[width=\textwidth]{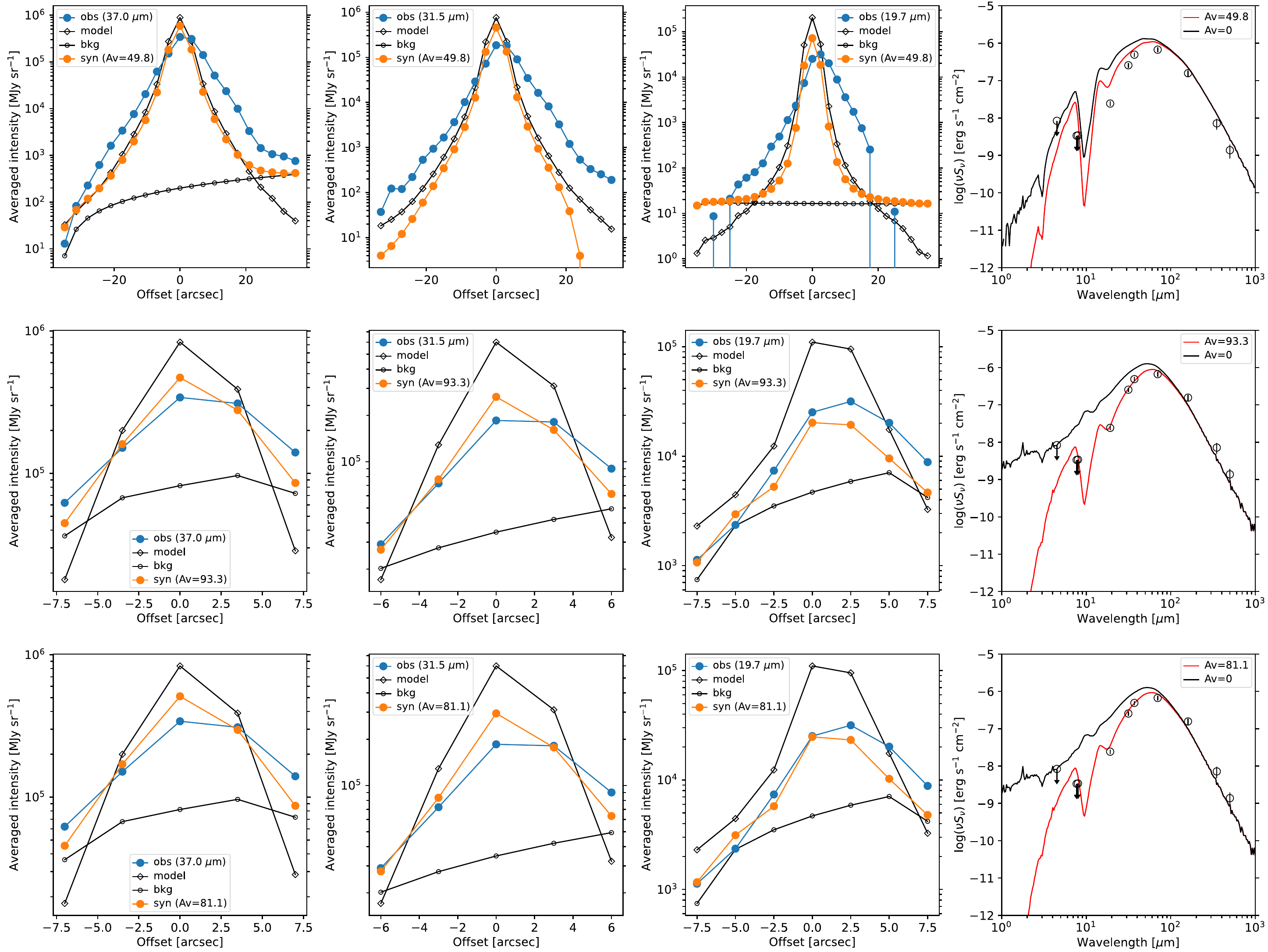}
  \caption{The best-fitting models of G35.2 fitted with the SED, IMPRO, and the combined SED+IMPRO (from top to bottom) among the models with at least 5 data points.  The SED values are extracted from a 37\farcs{5} aperture by \citet{2023ApJ...942....7F}.}
\end{figure*}

\addtocounter{figure}{-1}

\begin{figure*}[htbp!]
  \centering
  \includegraphics[width=\textwidth]{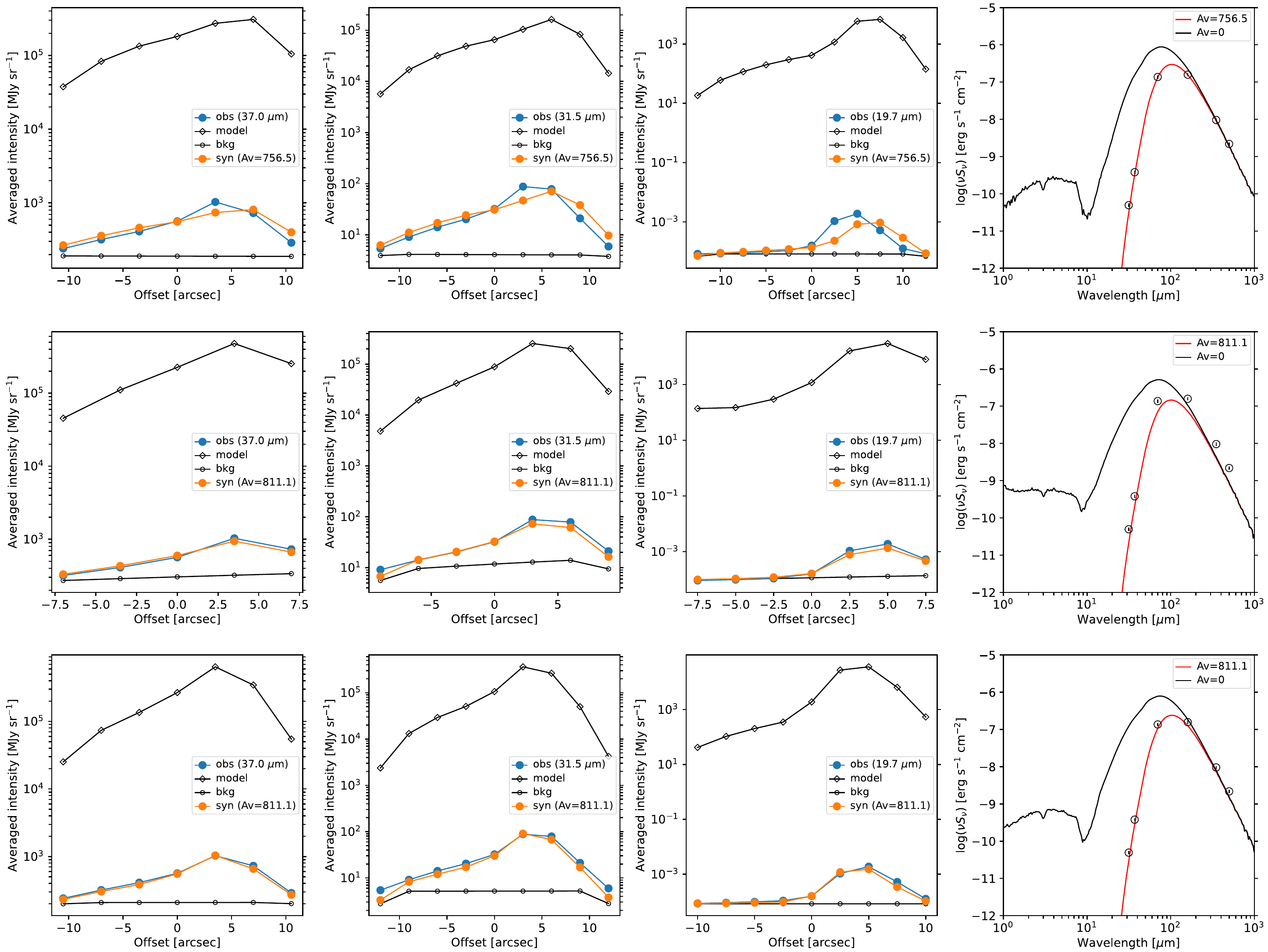}
  \caption{The best-fitting models of the synthetic ZT18 model fitted with the SED, IMPRO, and the combined SED+IMPRO (from top to bottom).  The SED values are extracted from the entire model, essentially adopting an aperture as large as the core size.}
\end{figure*}

\renewcommand{\thefigure}{\arabic{figure}}

\begin{figure}
    \centering
    \includegraphics[width=0.23\textwidth]{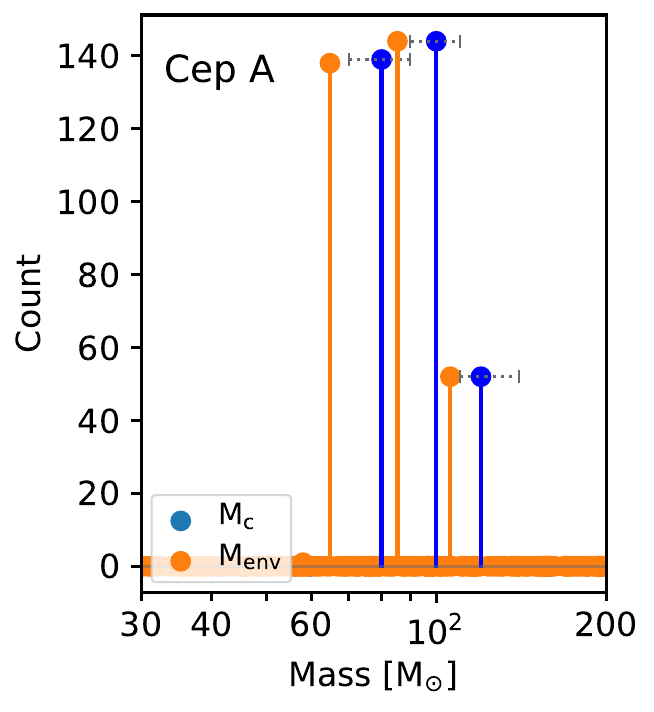}
    \includegraphics[width=0.23\textwidth]{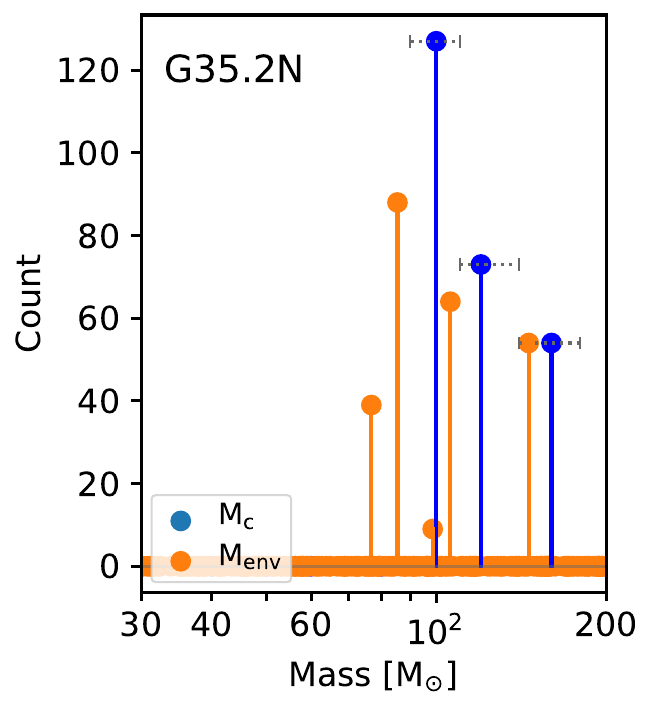}
    \caption{The envelope mass (\menv) of the good models for Cep A and G35.2N compared with the initial core mass (\mcore) shown in Figures\,\ref{fig:model_parameters} and \ref{fig:best_fitting_models}, respectively.  The \mcore\ is shown in the same style as that in Figure\,\ref{fig:model_parameters}, while the distribution of \menv\ is shown without the horizontal bar.}
    \label{fig:current_mass}
\end{figure}

\subsection{G35.20--0.74N - a case study for distant sources}
\label{sec:g35}

While the combined fitting pipeline estimates model parameters for Cep A comparable with observationally determined properties, we tested the limitation of this SED + IMPRO fitting approach using another well-known but more distant massive protostellar core, G35.20$-$0.74N \citep[hereafter G35.2N; e.g.,][]{2003MNRAS.339.1011G,2013AA...552L..10S,2014AA...569A..11S}.  The larger distance allows us to see how this fitting method performs with a worse physical resolution.  \citet{2013ApJ...767...58Z} presented a detailed modeling study using a similar theoretical framework as the one adopted here, where a single collapsing core could reproduce the SED and brightness profile along the outflow axis.  The success of this modeling work makes G35.2N an important target to benchmark our combined modeling pipeline.

G35.2N is a massive star-forming region consisting of 17 compact continuum sources in sub-millimeter wavelengths arranged in a filament-like orientation \citep{2014AA...569A..11S,2016AA...593A..49B,2022ApJ...936...68Z}.  Core A in \citet{2014AA...569A..11S}, also called Source 1 in \citet{2022ApJ...936...68Z}, has the highest total sub-mm flux of the entire region, while Core B (Source 2 in \citealt{2022ApJ...936...68Z}) has the second highest sub-mm flux of $\sim80\%$ of Core A.  These two sources dominate the entire region and drive their own outflows.  At sub-mm wavelengths, the outflows of Core A and B have position angles of $\sim30^\circ$ and $\sim6^\circ$, respectively \citep{2022ApJ...936...68Z}.  The outflows of Core A have a wide opening and extend to 40\arcsec--60\arcsec\ \citep[88,000--132,000 au;][]{2006AA...458..181B}, while the outflows of Core B are collimated and extend to $\sim20\arcsec$ \citep{2013ApJ...767...58Z}.  Outflow signatures are also discovered in near-infrared, mm, and radio wavelengths \citep{2003MNRAS.339.1011G,2006AA...458..181B,2015AA...573A..82C,2019NatCo..10.3630F}.  The separation between Core A and B is only $\sim2\arcsec$, which cannot be distinguished by the SOFIA observations.  To determine the dominant source at infrared wavelengths, we rely on archival images from 3.6 to 37 \micron, showing that the outflow from Core B dominates the infrared emission \citep{2006ApJ...642L..57D,2013ApJ...767...58Z}. Thus, for the fitting, we focus on Core B as the strip center.  We also adopt the outflow position angle of Core B, 6$^\circ$ \citep{2022ApJ...936...68Z}, for the orientation of the strip.

We modeled the SED and the multi-band 1D profiles of G35.2N with the same prescription as that for Cep A.  The only difference is the strip width.  A nominal 20\arcsec\ width at the distance of G35.2N, 2.2 kpc \citep{2009ApJ...693..419Z,2014AA...566A..17W}, corresponds to 44,000 au, which is larger than some core models.  More importantly, the strip width is chosen to best characterize the brightness profile along the outflow; therefore, the width for the G35.2N fitting should aim to have a similar physical size as the strip for Cep A.  Considering the angular resolution of SOFIA at $\sim20-40$ \micron\ ($\sim3.5\arcsec$), we chose a 10\arcsec\ width strip for extracting the 1D profiles to have adequate sensitivity and sufficiently narrow strip to probe the emission of the outflow.  The distributions of \chised, \chioned, and \chicomb\ are shown in Figures\,\ref{fig:chi2_sed_summary}, \ref{fig:chi2_1d_summary}, and \ref{fig:chi2_comb_summary}, respectively.

At the distance of G35.2N, the core radius of some models becomes smaller than twice the bin size of the 1D profile, determined by the beam size of SOFIA.  A 3\farcs{5} beam at 2.2 kpc corresponds to 7,700 au.  The ZT18 model grid has models with core radii as small as 2,700 au.  For all models with a core radius smaller than 7,700 au, we would only have one data point in the rebinned 1D profile, making the fitting ineffective.  Thus, we tested two selection criteria to only focus on the models that have at least three or five data points in all of their rebinned 1D profiles, essentially excluding the models with the core radii smaller than 7,700 au or 15,400 au.  Figure\,\ref{fig:model_parameters} show the parameter distributions of the good models, chosen to have $X^2_\text{min} \leq X^2 \leq X^2_\text{min} + 2$.  While the parameter distributions only change slightly for the SED fitting, the numbers of data points in the 1D profiles have a significant impact on the parameter distributions for the IMPRO fitting as well as the combined fitting.  The distributions become a lot tighter when the minimum number of data points increases from 3 to 5.  However, the corresponding $X^2_\text{min}$ becomes larger for the case with a minimum of 5 data points, suggesting that the fitting itself becomes worse.  Thus, increasing the minimum data points from 3 to 5 has no decisive improvement to the fitting.  For the following discussion on the fitting results of G35.2N, we focus on the fitting using a minimum of 3 data points in the 1D profiles.  The mean values and their corresponding ranges are listed in Table\,\ref{tbl:best_params}.  We also show the best-fitting models using SED, IMPRO, and the combined SED+IMPRO with at least 3 and 5 data points in Figure\,\ref{fig:best_fitting_models}.

The SED fitting and the 1D profile fitting produce different mean values of \mcore\ by $\sim100$ \msun\ and both distributions show an irregular shape with a gap, suggesting no obvious best-fitting value.  By combining the SED and IMPRO fitting, the \mcore\ distribution becomes a single coherent distribution, improving the constraining power of the fitting.  The envelope mass (\menv) of the good models provides a realistic comparison with observational constraints.  Figure\,\ref{fig:current_mass} (right) shows the \menv\ distribution for G35.2N.  These models have a core radius of $\sim$12000 au.  In comparison, \citet{2009AA...499..811L} derived a clump mass of 110--250 \msun, using C$^{18}$O emission that traces the dense gas.  Their observations were taken by IRAM 30m telescope with an angular resolution of 11\arcsec, which is comparable to the estimated core size.  From the IMPRO+SED fitting, the envelope and initial core mass have geometric mean values of 99.8 and 116.5 \msun, respectively.  The \menv\ is lower than the observationally constrained clump mass, which is expected that the G35.2N is known to have multiple cores \citep{2022ApJ...936...68Z} and our fitting method only focuses on the dominant core in the system.  The estimated \menv\ is $\sim10\%$ lower than the lower limit of the measured mass, confirming the assumption of a singly dominant core.

The \sigmacl\ from SED fitting has a primary peak at 0.4 g cm$^{-2}$ with a broad distribution covering the entire sampling range (0.1--3~g cm$^{-2}$), while the \sigmacl\ becomes 1.9$^{+1.3}_{-0.9}$ g cm$^{-2}$ from the IMPRO fitting.  In the combined fitting, all the good models have the \sigmacl\ of 3.2 g cm$^{-2}$, which is also where the majority of the good models fitted with IMPRO.  For \mstar, the parameter distributions shift toward lower values from the SED to the IMPRO fitting.  The combined fitting has a similar distribution as that in the IMPRO with a narrower spread.  The distribution of \view\ changes substantially between SED and IMPRO fitting.  With the IMPRO, the distribution has a prominent peak at $\sim35^\circ$ and a secondary peak at $\sim65^\circ$.  The distribution of \view\ in the combined fitting also shows a bimodal feature, but the secondary peak is much more suppressed compared to that in the IMPRO fitting.  For \av, both SED and IMPRO fitting fail to constrain the value to a narrow range.  The IMPRO fitting limits the good models to a few models, but the overall span of the distribution remains broad.  In the combined fitting, the value of \av, as expected from the lack of desired performance in the SED and IMPRO fitting, is not well constrained. In general, the improvement from the inclusion of 1D profile fitting is not apparent in the case of G35.2N, suggesting limitations in this fitting method in poorly-resolved sources (see Section\,\ref{sec:limitation}).

Figure\,\ref{fig:model_parameters} shows the parameter distributions using at least 5 data points in the IMPRO fitting.  By setting a lower limit in the core radius, there are fewer models considered in the selection of good models.  All the resulting distributions become singly peaked in contrast with several bimodal distributions in the results with at least 3 data points, such as \view\ and \av.  Compared to the best-fitting values with at least 3 data points, the \sigmacl\ becomes lower while the \mstar, \view, and \av\ become higher.  The best-fitting \view\ with at least 5 data points is consistent with the secondary peak seen in the distribution with at least 3 data points. As there is no robust measurement of the core radius, we cannot put a rigorous limit on the minimum number of data points.  Thus, we adopt the results fitted with at least 3 data points to avoid selecting the small cores that cannot be distinguished by the IMPRO method and preserve the maximum freedom in the fitting.

For Core B, \citet{2013AA...552L..10S} modeled the velocity structure of \methylcyanide\ with Keplerian rotation and estimated a central mass of 18$\pm$3 \msun, which also includes a nearby continuum source separated by 0\farcs{38} (see also \citealt{2016AA...593A..49B}).  This companion source likely contributes little to the estimated mass because its dust mass is 20 times lower than that of Core B.  Recently, in Core B, \citet{2022ApJ...936...68Z} modeled the free-free emission with a simple disk model and suggested a ZAMS stellar mass of 11.7 to 18.4 \msun.  They also discovered a spiral structure traced by SO$_2$ emission, indicative of instability of a massive disk.  The best-fitting \mstar\ of 8.6 \msun\ is consistent with the mass of Core B.  \citet{2013ApJ...767...58Z} adopted a similar turbulent core model, where a clump is added and the outflows extend further outside the core radius, and found a \mstar\ of 22--34 \msun.  While this value is higher than the best-fitting value using at least 3 data points, it is consistent with the result using at least 5 data points, \mstar=27.3$^{+4.7}_{-3.3}$ \msun.  Interestingly, their best-fitting core radius, 11400 au, is 50\%\ of that in our best-fitting model.  However, they include a clump in addition to the core model explored in our fitting pipelines.  Thus, the effective core size may be larger than 11400 au, making it closer to the core radii suggested by our fitting approaches.

\citet{2015AA...573A..82C} derived \av\ in the G35.2N region, but they cannot constrain the \av\ toward Core B (knot 2 in their study) due to lack of sufficient [Fe~\textsc{ii}] lines detection.  Toward a nearby Core A, they derive an \av\ of 24$\pm$4 mag.  \citet{2022ApJ...936...68Z} show a stronger sub-mm continuum emission toward Core A compared to Core B, which may suggest the \av\ toward Core B could be lower than that toward Core A.  However, deriving \av\ from sub-mm interferometric observations is confused by the limited maximum recoverable scale.  Using IMPRO+SED, we constrain an \av\ of 34.2$^{+52.8}_{-32.3}$ with a bimodal distribution, qualitatively consistent with the \av\ derived by other studies. If we select the models with at least 5 data points, the \av\ becomes narrowly constrained to 85.9$^{+21.3}_{-24.6}$, resulting in a larger tension with literature values.

The rotating infalling envelope model constrained by the SO$_2$ emission suggests a disk inclination of 30$^\circ$ with respect to the line of sight \citep{2022ApJ...936...68Z}, corresponding to a \view\ of 60$^\circ$ if the outflow is perpendicular to the disk midplane.  Our combined fitting suggests a \view\ of 30.3\arcdeg$^{+3.7\arcdeg}_{-8.3\arcdeg}$.  The best-fitting \view\ is lower than the inclination inferred from the disk midplane, which may suggest misalignment between the outflow and the disk.  The outflow-disk misalignment is expected in the multiple systems formed in a turbulent environment \citep{2016ApJ...827L..11O}, which is consistent with the turbulent core modeling framework adopted in this study.  Interestingly, the best-fitting value using at least 5 data points, 60.5$^{+7.5}_{-9.5}$, agrees with the \view\ inferred from the disk orientation.  The notable differences in best-fitting values due to the limit on the minimum data points indicate that the fitting results are less reliable compared to those of Cep A.  Independent constraints on the minimum core size, for example, obtained by higher resolution observations, would be valuable to improve the fitting accuracy.

\subsection{Limitations of combined SED and IMPRO Fitting}
\label{sec:limitation}

\subsubsection{Degeneracies in IMPRO and SED Fittings}

The fitting methods developed and tested in this study aim to provide a practical approach to find the best-fitting model parameters that reproduce observations, under the model framework constructed by ZT18.  While the SED+IMPRO fitting provides mostly consistent estimates of the physical properties in the benchmarking tests on Cep A and G35.2N as well as the retrieval test from a synthetic ZT18 model, we also discover limitations and degeneracies in the fitting results from the $\chi^2$ distributions.  We discuss the limitations in the following paragraphs, separating by the fitting methods (SED, IMPRO, and SED+IMPRO).  

In \chised, we observe wedge-like distributions in \mstar-\view\ and \mcore-\view.  \mstar\ is constrained to a range of values but not narrowly constrained, whereas the fitting only provides lower limits on \view.  In the retrieval test, the \view\ can still be constrained with a lower contrast in the \chised\ value despite that similar wedge-like patterns still appear in \mstar-\view\ and \mcore-\view.  Both \mstar\ and \mcore\ are constrained moderately with \mcore\ tending to have a larger range of best-fitting values compared to that of \mstar.  The distribution of \mcore-\sigmacl\ shows a clear anti-correlation pattern, suggesting a degeneracy.  For \sigmacl, its minimum \chised\ corresponds to a narrow range of \mstar, \view, and \av.  Lastly, in \av, the \chised\ can only indicate either lower or upper limits.

In \chioned, the distributions of \mstar-\mcore, \mcore-\sigmacl, and \mcore-\view\ all have clear coherent minima, suggesting tighter constraints.  We observe a fan-like pattern in \view-\av, indicating the presence of multiple correlated best-fitting values; however, a localized range of better fitting \view\ can be seen in Figure\,\ref{fig:chi2_1d_summary}.  The IMPRO fitting has less constraint on \av, although the effect of \av\ is included because measurements at multiple wavelengths are included.  Particularly, in the distribution of \mstar-\av, the \mstar\ is moderately constrained whereas \av\ still has a broad range of values.  

Both the SED and IMPRO methods provide consistent estimates on \mcore\ and \mstar.  However, the \sigmacl\ estimate varies significantly because of the difference in \view\ and \av.  IMPRO constrains \view\ much better.  Since the \sigmacl\ with minimum \chisq\ connects to a narrow range of \mstar, \view, and \av, a better constrained \view\ and the varying prediction of \av\ can significantly affect the estimated \sigmacl.  For \av, neither fitting method alone provides a reliable and consistent estimate.  

The combination of SED and IMPRO fitting aims to take the advantage of their asymmetric constraining power in the model parameters.  The IMPRO fitting is designed to be sensitive to \view\ and the distribution of \chioned\ confirms its sensitivity (Figure\,\ref{fig:chi2_1d_summary}).  With the SED+IMPRO, the constraint on \view\ is similar to that observed in the IMPRO fitting.  The distribution of \mcore-\sigmacl\ still shows anti-correlation; however, the \sigmacl\ can be better localized because of the improved constraint on \view.  Thus, while \mcore\ can be moderately constrained with either the SED (e.g., Cep A and synthetic ZT18 model) or IMPRO (e.g., G35.2N), the combined SED+IMPRO fitting provides a better estimate on \mcore.  In the combined fitting, \av\ remains as the least constrained parameter among the five model parameters. 

In summary, compared to the SED fitting, the IMPRO fitting provides improved constraints on \view, which are benchmarked with independent observational constraints as well as a retrieval test using synthetic ZT18 model.  With a better constrained \view, the estimate \sigmacl\ is also improved.  Among the model parameters in the ZT18 model grid, \av\ remains as the least constrained parameter, while the core-specific properties, \mcore, \sigmacl, \mstar, and \view\ can be estimated.  A caveat of the current work is that the IMPRO + SED fitting has only been carried out on a limited number of sources. Future papers in this series will expand the sample of sources that are explored by this method.

\subsubsection{Spatial Resolution}
\begin{figure*}[htbp!]
  \centering
  \includegraphics[height=1.4in]{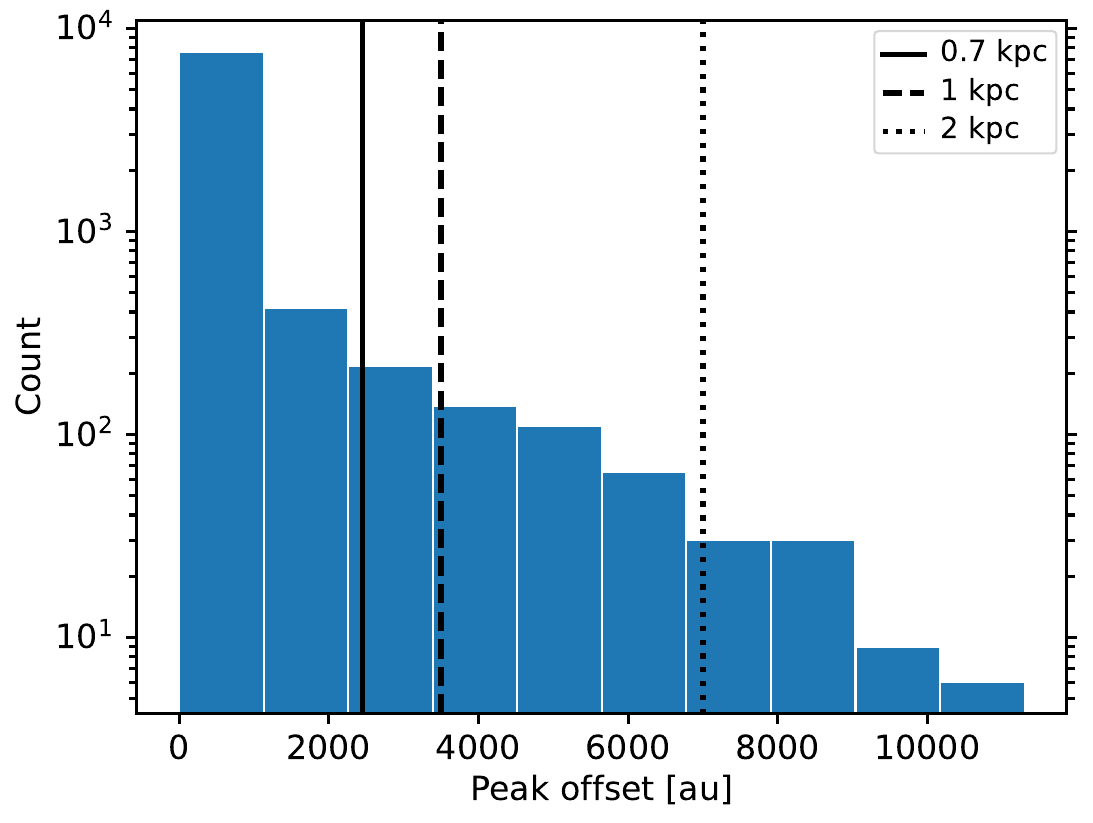}
  \includegraphics[height=1.5in]{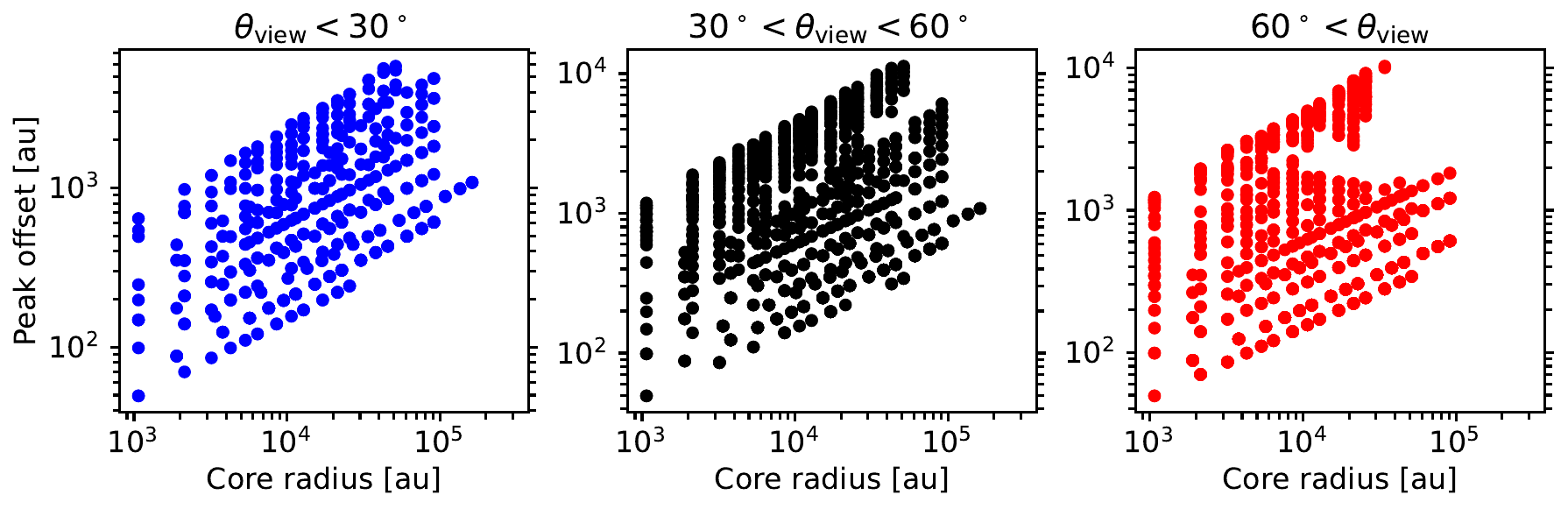}
  \caption{Left: The distribution of the offsets of the peak brightness derived from the modeled images using a 20\arcsec\ width strip.  The solid, dashed, and dotted lines indicate the physical resolution of SOFIA at 37 \micron\ at 0.7, 1.0, and 2.0 kpc.  Right: Peak offset compared with core radius in the ZT18 model grid.  The models are separated into three panels based on their \view.}
  \label{fig:separation}
\end{figure*}

As we demonstrated in the case of G35.2N (Section \ref{sec:g35}), combining IMPRO 1D intensity profile along the outflow axis fitting with SED fitting has little impact on improving constraints on ZT18 model grid parameters.  On the other hand, in Cep A, after the inclusion of the 1D profiles, the parameter distributions of the good models substantially tighten. The main reason for the drastic difference in performance is likely the linear resolution of the observations.  The distance of Cep A is 0.7 kpc, compared to 2.2 kpc for G35.2N.  Thus, with SOFIA-FORCAST, the linear resolution for Cep A and G35.2N is $\sim2500$ au and $\sim7700$ au, respectively.  As a result of the three times lower linear resolution for G35.2N, we needed to limit model testing for those with at least three data points in the 1D profiles.  If the brightness profile along the outflow cannot be sufficiently resolved, adding the 1D profile to the SED fitting naturally has little effect.
Figure\,\ref{fig:separation} shows the distribution of the peak offsets in the 1D profiles for all the models in the ZT18 grid. A majority of the models have peak offsets $\lesssim$2000 au, which is marginally resolved in the case of Cep A.  At the distance of G35.2N, only a minor subset of models can be distinguished.  The contrast and magnitude of the asymmetry in the brightness profiles also depend on the core properties.  Smaller cores tend to have smaller peak offsets in their outflows regardless of their \view\ (Figure\,\ref{fig:separation}, three panels on the right).
Thus, while IMPRO fitting can provide unique constraints to the model parameters, observations with sufficient spatial resolution are required.

\subsubsection{Multiplicity}
Intrinsic source properties may also affect the performance of the profile modeling.  Our modeling approach implicitly assumes a single protostar dominating the observed core, which has a symmetric outflow.  However, massive star-forming regions can have multiple sources.  G35.2N has at least four sources, each driving an outflow.  While we focused on the likely dominant source for modeling, the presence of other sources may still contribute significantly to the observed brightness morphology and the SED.  Furthermore, the bipolar outflow may not be symmetric as assumed in the model.  Asymmetry in outflow cavities and jets has been seen in multiple low-mass protostellar outflow \citep{2021ApJ...911..153H,2023ApJ...947...25H}.  Such asymmetries can occur in high-mass outflows as well \citep[e.g.,][]{2023AA...676A.107F,2025ApJ...983...19C}.  Higher resolution observations will also help us better constrain these effects to revise the models of massive protostellar cores, as well as the fitting approach.

\section{Conclusions}
\label{sec:conclusions}

In this study, we developed an Image Profile (IMPRO) fitting method based on MIR/FIR intensity profiles along the outflow axis to constrain the fundamental properties of massive protostars. This IMPRO fitting has also been combined with SED fitting to yield the best joint constraints. In particular, for IMPRO fitting, we utilize the 1D intensity profile extracted from SOFIA-FORCAST images taken as part of the SOFIA Massive (SOMA) Star Formation survey at 19.7, 31.5, and 37.1 \micron.  Using Cep A as a test case, we found that the combined SED and IMPRO fitting approach provides better constraints on protostellar parameters in the ZT18 model grid.  Particularly, the viewing angle (\view) becomes tightly constrained with the addition of the 1D intensity profiles.  However, for G35.2N, a more distant source, this combined fitting method yields little improvement compared to the SED fitting alone due to the limited spatial resolution of the SOFIA-FORCAST observations. Thus, when fitting massive protostellar cores at distances $\sim 2\:$kpc or greater, higher resolution images are required to take advantage of the constraining power from the 1D profiles. We expect this fitting method to be applied to higher resolution infrared images taken by JWST NIRCam and MIRI.

\begin{acknowledgements}
Y.-L.Y. acknowledges the support from the Virginia Initiative of Cosmic Origins (VICO) Postdoctoral Fellowship. JCT acknowledges support from ERC Advanced Grant 788829 (MSTAR) and NSF grant AST–2206437. R.F. acknowledges support from the grants PID2023-146295NB-I00, and from the Severo Ochoa grant CEX2021-001131-S funded by MCIN/AEI/ 10.13039/501100011033 and by ``European Union NextGenerationEU/PRTR''.  Based on observations made with the NASA/DLR Stratospheric Observatory for Infrared Astronomy (SOFIA) in the Early Science Period and program 02\_0074. SOFIA is jointly operated by the Universities Space Research Association, Inc. (USRA), under NASA contract NNA17BF53C, and the Deutsches SOFIA Institut (DSI) under DLR contract 50 OK 0901 to the University of Stuttgart. This research has made use of the NASA/IPAC Infrared Science Archive, which is funded by the National Aeronautics and Space Administration and operated by the California Institute of Technology.
\end{acknowledgements}

\facilities{\textit{SOFIA}}

\software{Astropy \citep{2013AA...558A..33A,2018AJ....156..123A,2022ApJ...935..167A}}

\appendix

\section{Effect of viewing angle}
\label{sec:syn_view}

\begin{figure}[htbp!]
    \centering
    \includegraphics[width=\linewidth]{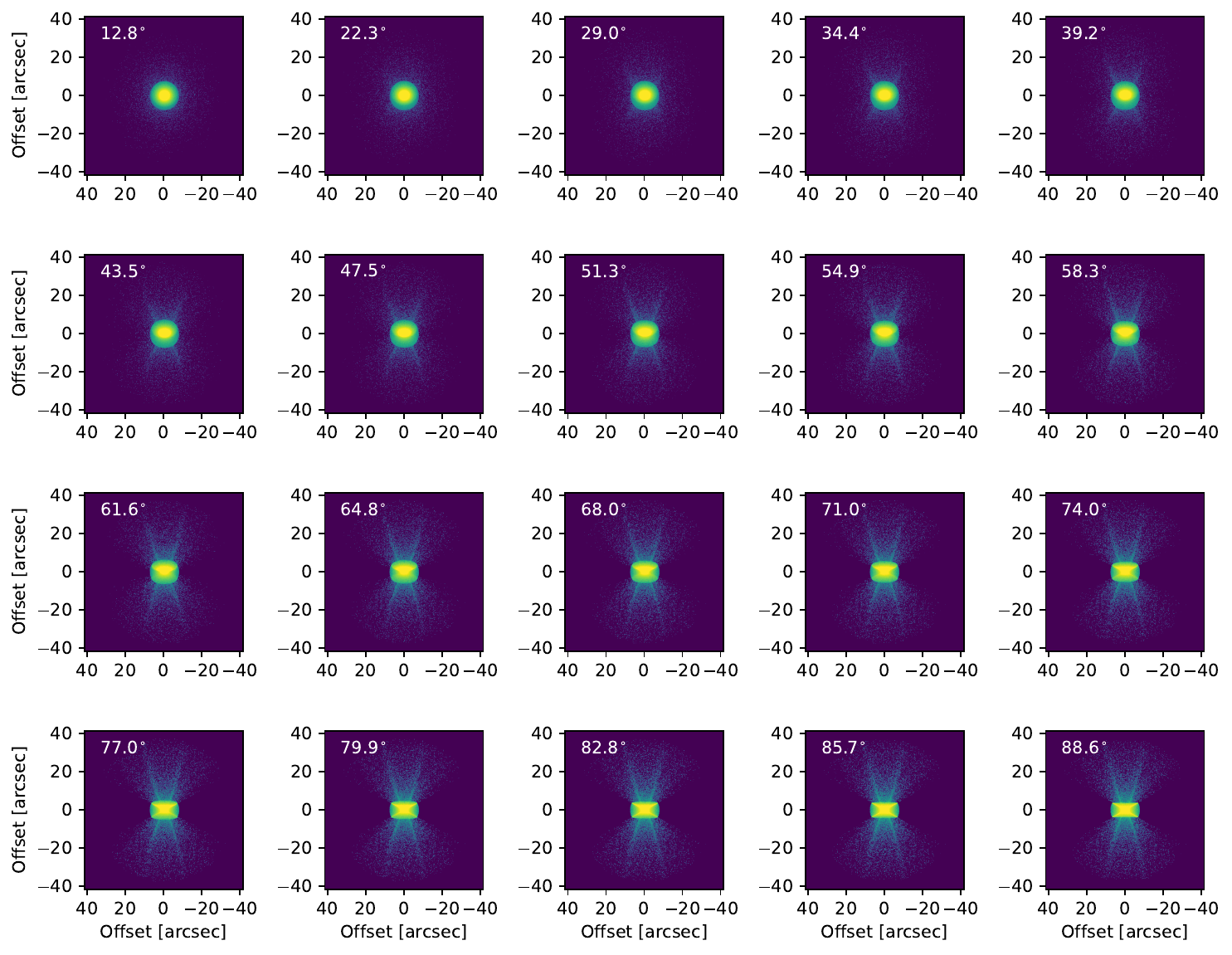}
    \caption{Synthetic 37.1 \micron\ images of a fiducial model at the \view\ sampled by the ZT18 model grid.  The \view\ is labelled in each panel.}
    \label{fig:syn_view_grid}
\end{figure}

\begin{figure}[htbp!]
    \centering
    \includegraphics[width=0.32\linewidth]{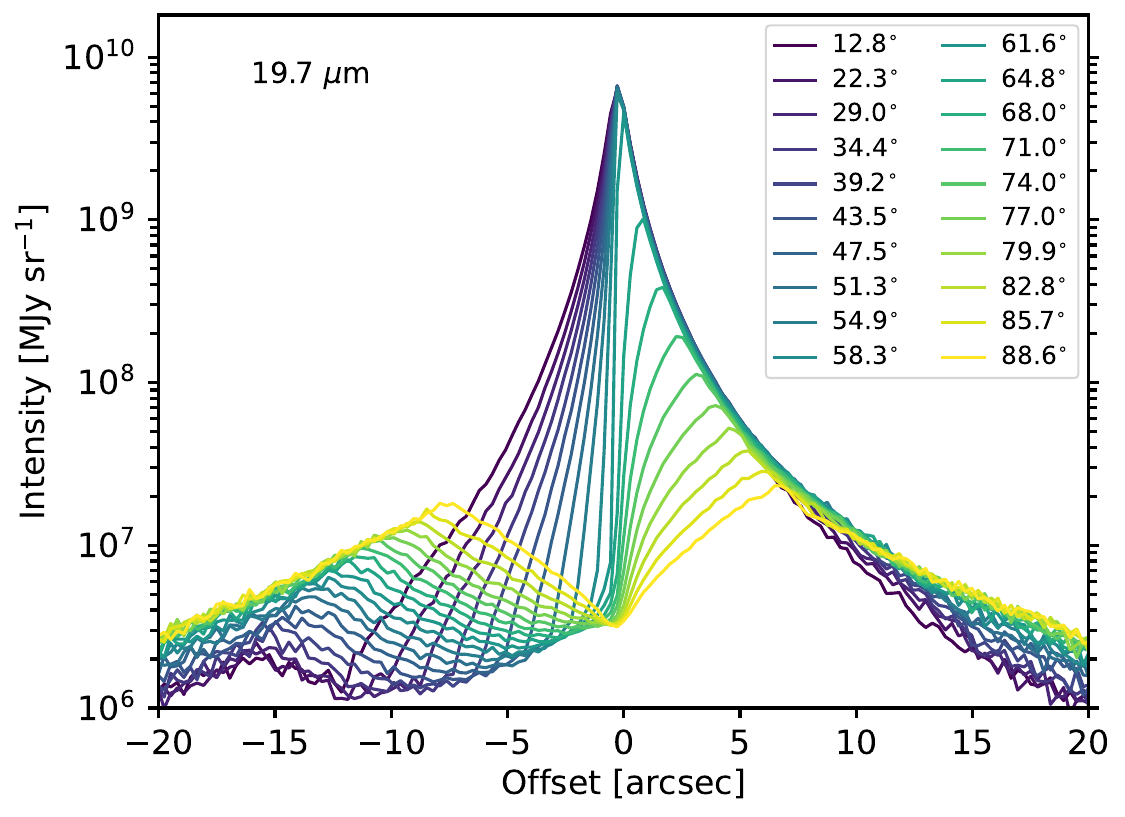}
    \includegraphics[width=0.32\linewidth]{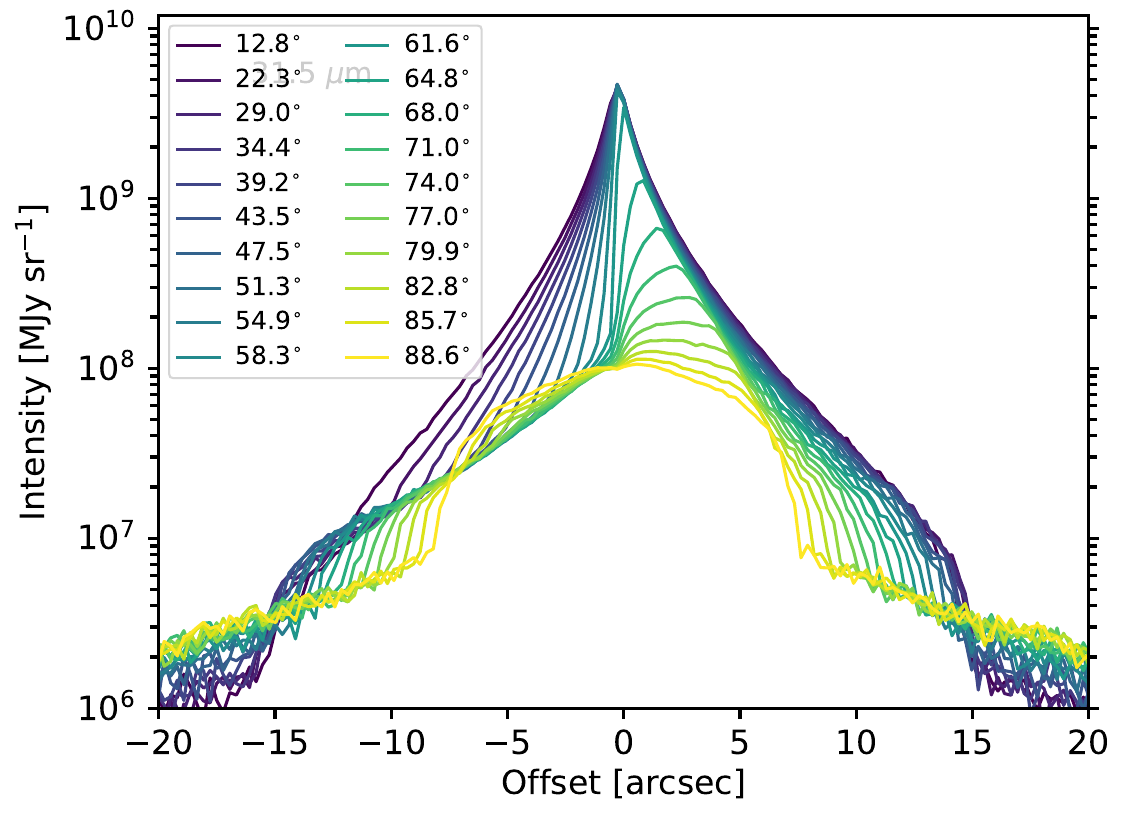}
    \includegraphics[width=0.32\linewidth]{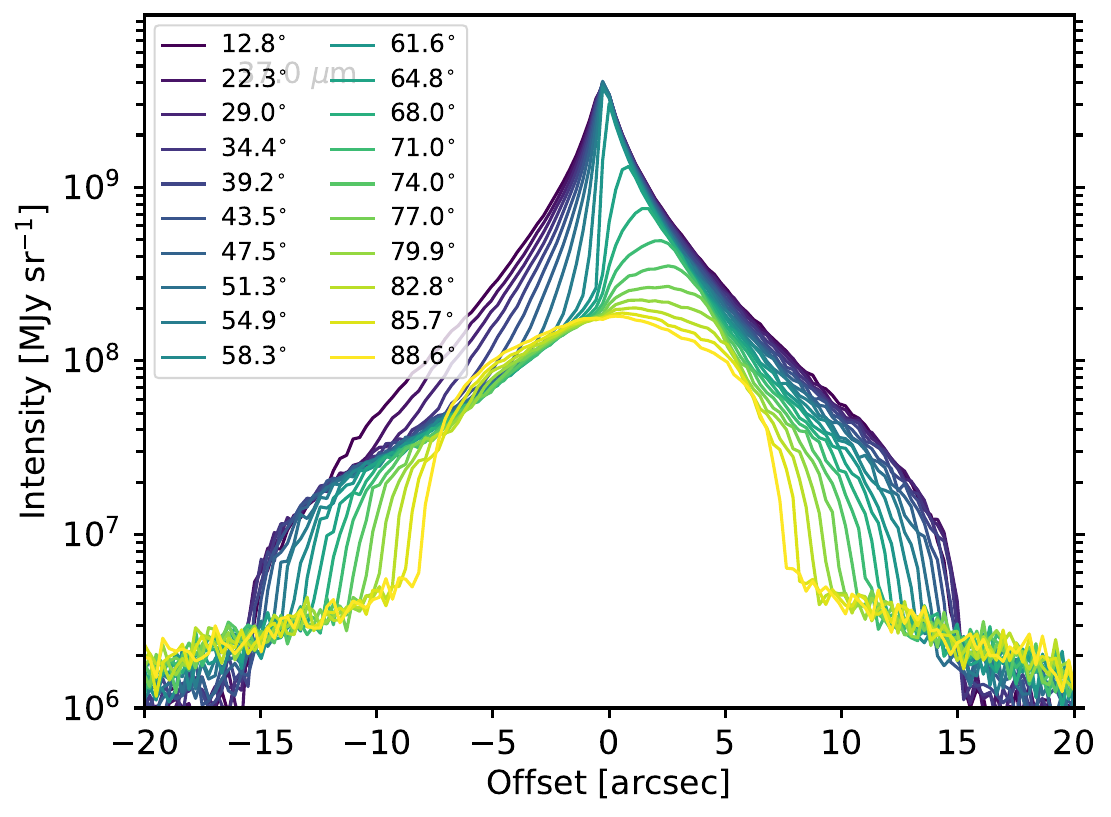}
    \includegraphics[width=0.32\linewidth]{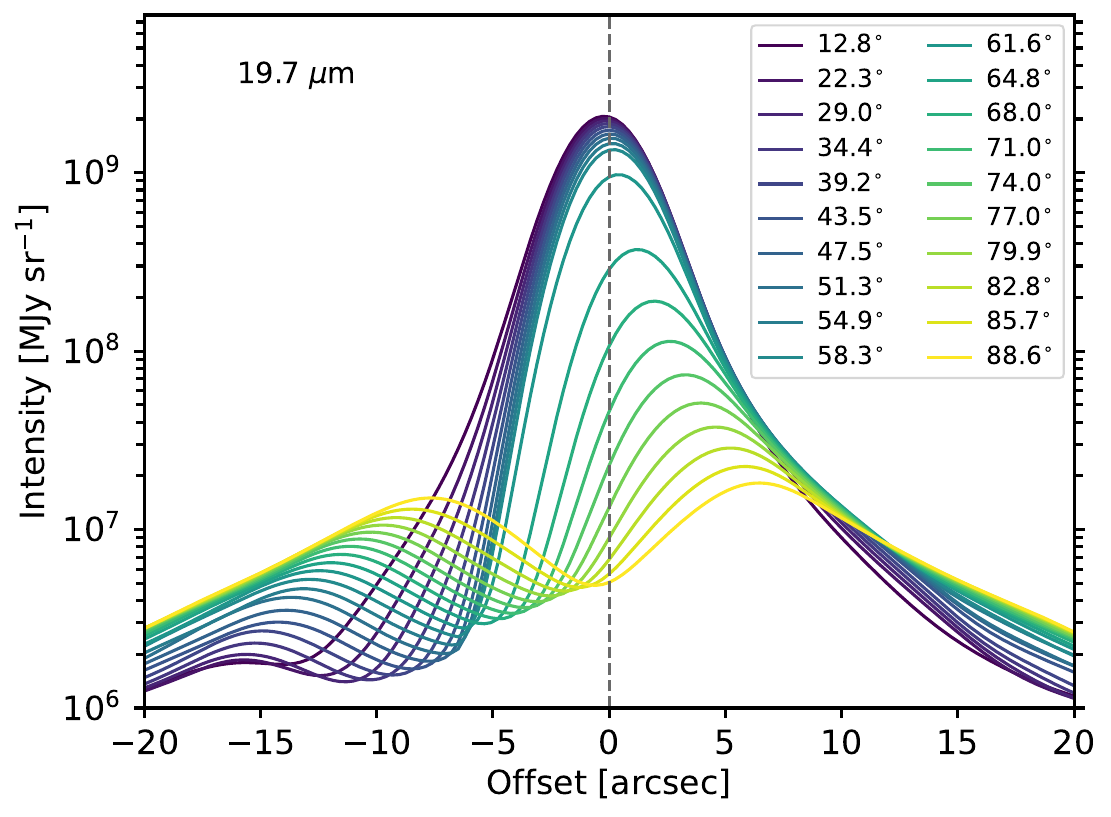}
    \includegraphics[width=0.32\linewidth]{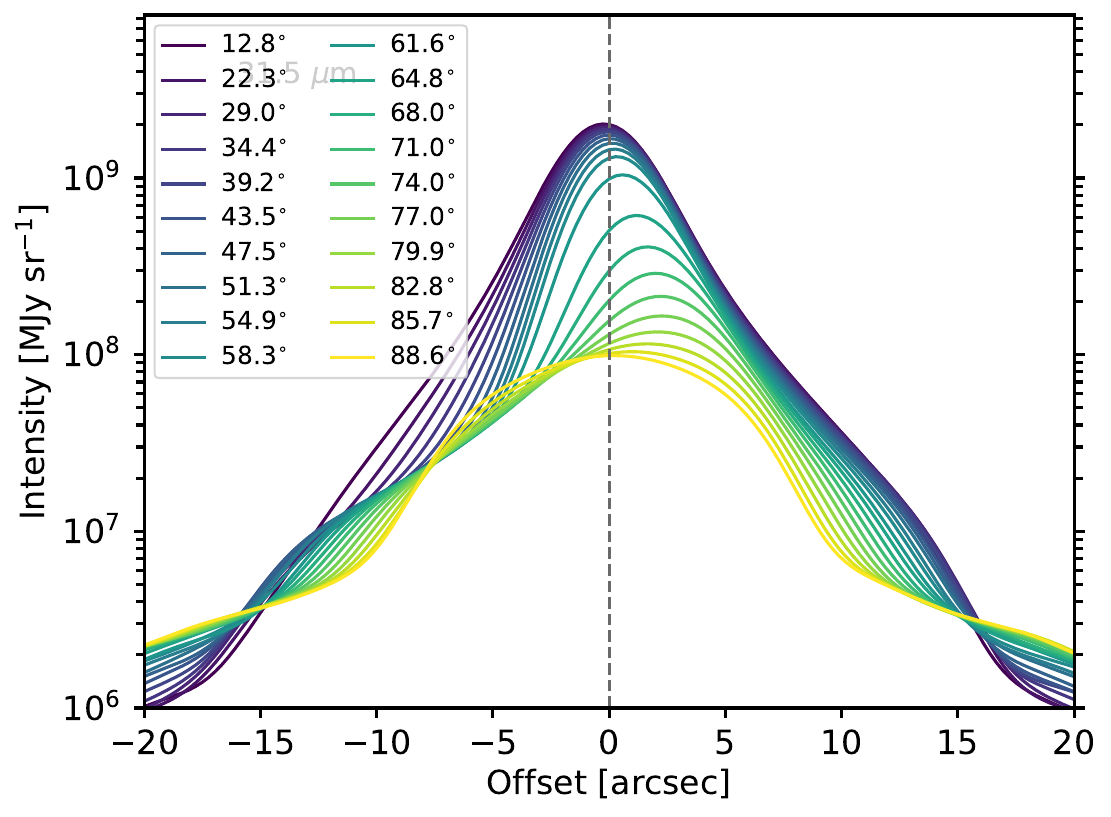}
    \includegraphics[width=0.32\linewidth]{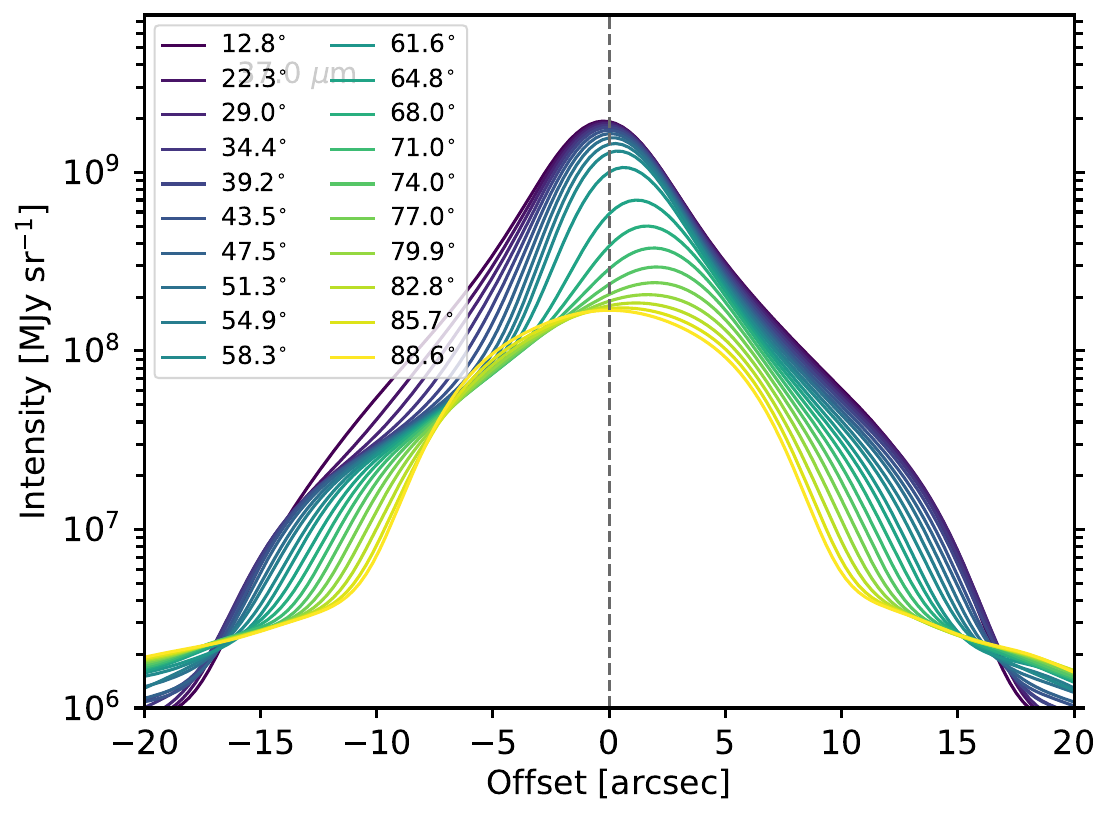}
    \caption{Synthetic 1D brightness profile along the outflows in the ZT18 model grid at the sampled \view.  The top and bottom rows show the original profiles and the profiles smoothed with corresponding beam sizes, respectively.  The beam sizes are 2.5\arcsec, 3.0\arcsec, and 3.5\arcsec\ at 19.7, 31.5, and 37.1 \micron\ (left to right), respectively.}
    \label{fig:syn_view}
\end{figure}

When outflows are viewed edge-on (\view=90$^\circ$) and pole-on (\view=0$^\circ$), the brightness along the outflow should be symmetric.  Indeed, the synthetic profiles are close to symmetric at the smallest and the largest \view, close to pole-on and edge-on (Figure\,\ref{fig:syn_view_grid}).  Figure\,\ref{fig:syn_view} shows the synthetic 1D brightness profiles as a function of \view\ in the ZT18 model grid.  The asymmetry is most significant at the shortest wavelength and at \view$\sim50^\circ-80^\circ$.  The contrast in the 1D brightness profile becomes weaker when the 1D profiles are smoothed to the corresponding beams, but the asymmetry is still apparent.  

\section{Benchmarking the fitting pipelines with the ZT18 grid}
\label{sec:benchmark}

We conducted a retrieval test using a synthetic model from the ZT18 model grid to benchmark the performance of the IMPRO and the SED+IMPRO pipelines.  Ideally, for a given synthetic observation, the fitting pipeline should return the exact same model.  However, the uncertainties added to simulate the uncertainties in observations, as well as the degeneracies in the model grid, could reduce the predictability of the pipelines.

We first randomly select a model, \zt{120}{3.16}{8.0}{34}, and an \av\ of 779.7, as the fiducial model.  We scaled the model images at 19.7, 31.5, and 37.1 \micron\ from the default 1 kpc to 0.7 kpc to study the performance at the distance of Cep A, simulating one of our test cases.  Then, we added a flat background with 10\%\ uncertainty, following the same assumption as our processing of the observed data.  The flat background is set to have \eten{3}, \eten{4}, and \eten{5} MJy sr$^{-1}$ at 19.7, 31.5, and 37.1 \micron, respectively, matching the background levels in the observations.  For the SED, we use \texttt{SedFitter().get\_sed()} from \texttt{sedcreator} to apply the distance and extinction.  We took all the fluxes in the synthetic SED, essentially adopting an aperture as large as the core size.  

We ran the same fitting pipelines (SED, IMPRO, and SED+IMPRO) using the synthetic images and SED.  Figure\,\ref{fig:model_parameters} shows the parameter distributions of the good models, while Figures\,\ref{fig:chi2_sed_summary}, \ref{fig:chi2_1d_summary}, and \ref{fig:chi2_comb_summary} show the $X^2$ behaviors in comparison with model parameters.  Figure\,\ref{fig:best_fitting_models} shows the best-fitting models using the fitting pipelines.  The SED+IMPRO fitting retrieves the model parameters within uncertainties.  Compared to the fitting of Cep A, the SED-only fitting constrains the parameters closer to the input parameters.  The \mcore\ shows a bimodal distribution with only a few models selected as good models.  The good \view\ has a large range.  Compared to the SED fitting, the IMPRO-only fitting overestimates the \mcore\ and underestimates the \sigmacl\ with a bimodal distribution.  Interestingly, the IMPRO produces a poorer constraint on \view\ but only a few models are selected in each bin.  In the SED+IMPRO fitting, all parameters are constrained and consistent with the input parameters.  The \mcore\ still shows bimodal distribution, but the range of the \mcore\ is narrower than that in the SED fitting, suggesting that the parameter is better estimated.  This test demonstrates that both SED and IMPRO are required to produce a satisfactory fitting result.

\section{Fitting uncertainty due to position angle}
\label{sec:pa_test}
\begin{figure*}
    \includegraphics[width=\textwidth]{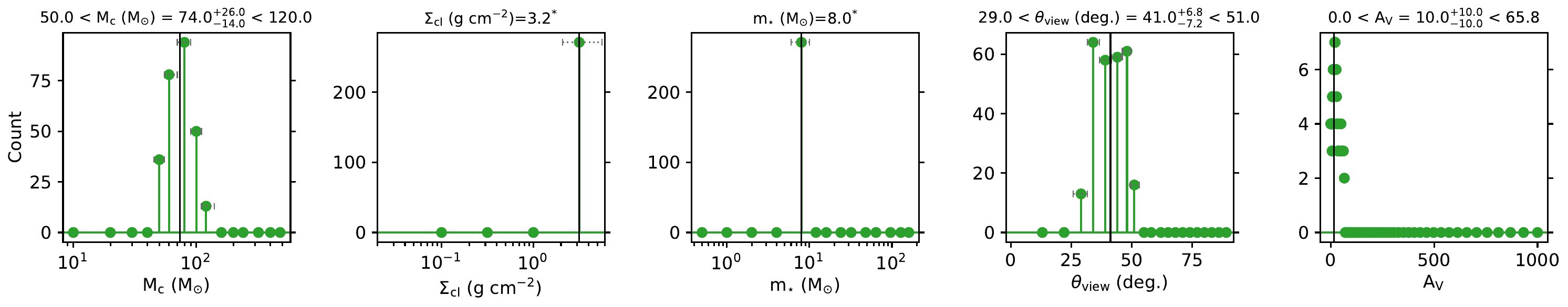}
    \includegraphics[width=\textwidth]{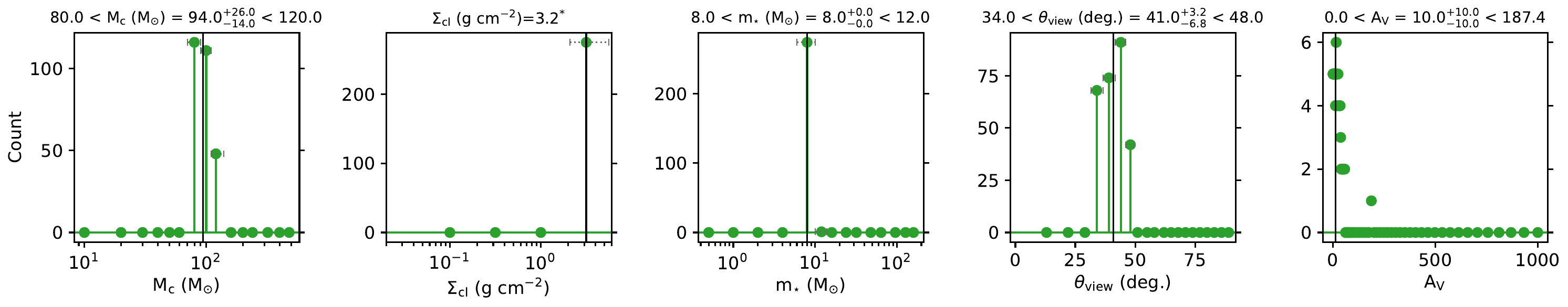}
    \includegraphics[width=\textwidth]{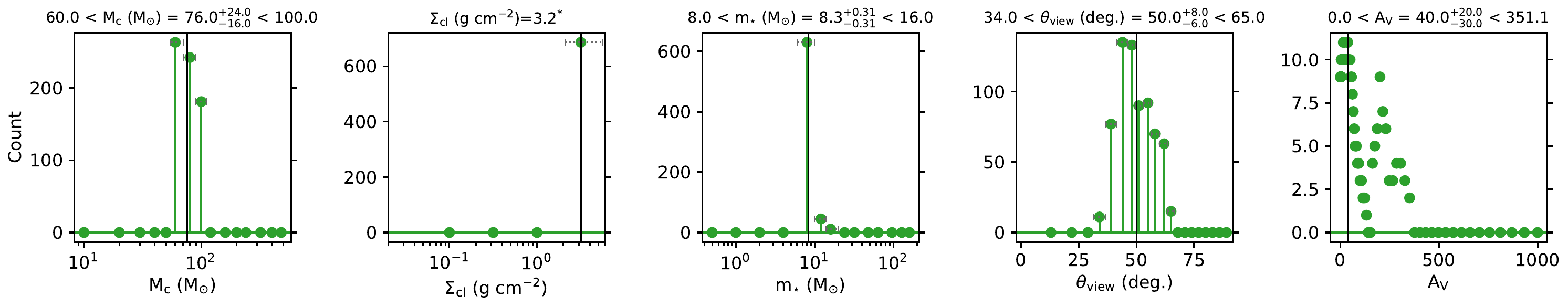}
    \caption{Distributions of the fitted parameters for the good model using PA of 70\arcdeg\ (top), 50\arcdeg\ (middle, the fiducial PA), and 30 \arcdeg\ (bottom).  Only the results fitted with the IMPRO pipeline are shown here to emphasize the uncertainty due to the choice of PA.  The minimum \chioned\ are 3.5, 5.6, and 8.4 for PA of 70\arcdeg\ (top), 50\arcdeg\ (middle), and 30 \arcdeg\ (bottom), respectively.  An asterisk indicates that all models have the same value.}
    \label{fig:pa_test}
\end{figure*}

We explore the uncertainties in the fitted parameters due to the choice of outflow position angle (PA).  We ran the same fitting procedures as those described in Section\,\ref{sec:cepA_all} but with +20\arcdeg\ and $-20$\arcdeg\ from the fiducial PA of 50\arcdeg.  Figure\,\ref{fig:pa_test} shows the distributions of the good model parameters using $\pm$20\arcdeg\ from the fiducial PA.  The \sigmacl\ and \mstar\ have no significant difference with a 20\arcdeg\ uncertainty in the PA, whereas the \mcore\ becomes slightly smaller in both +20\arcdeg\ and $-$20\arcdeg\ cases.  The \view\ also varies slightly ($<10\%$) with the PA, but the distributions remain consistent.  With a misaligned PA of $-$20\arcdeg, the \av\ distribution extends toward higher values, leading to a larger best-fitting value.  In comparison, a misaligned PA of +20\arcdeg\ produces a similar \av\ parameter distribution.  In summary, an uncertainty of 20\arcdeg\ in PA has only minor effects on the fitted parameter distributions.  Thus, unless the outflow has a very wide opening angle and/or a complex structure, outflow PA constrained by molecular emission observed with sufficient spatial resolution should provide suitable PA constraints for the IMPRO fitting approach.

\end{document}